\documentclass[a4paper,11pt]{article}
\usepackage{a4wide}
\usepackage{amsmath,amsfonts,bbm,mathrsfs,amsthm}
\usepackage[dvips]{graphics,graphicx,psfrag}
\usepackage{epsfig}
\usepackage{color}

\allowdisplaybreaks

\psfrag{1}{$1$}
\psfrag{2}{$2$}    
\psfrag{3}{$3$} 
\psfrag{ps11a}{$\Psi_1^{(11)}$}
\psfrag{ps11b}{$\Psi_2^{(11)}$}
\psfrag{ps12}{$\Psi^{(12)}$}
\psfrag{ps12a}{$\Psi^{(12)}_1$}
\psfrag{ps12b}{$\Psi^{(12)}_2$}
\psfrag{ps21}{$\Psi^{(21)}$}
\psfrag{ps21a}{$\Psi^{(21)}_1$}
\psfrag{ps21b}{$\Psi^{(21)}_2$}
\psfrag{ps22}{$\Psi^{(22)}$}
\psfrag{ps22a}{$\Psi^{(22)}_1$}
\psfrag{ps22b}{$\Psi^{(22)}_2$}

\newcommand{\mat}[4]{\left(\begin{array}{cc}#1&#2\\#3&#4\end{array}\right)}
\newcommand{\vm}{\vec{m}}
\newcommand{\vn}{\vec{n}}

\begin{document}
\thispagestyle{empty}
\begin{flushright}
MPP-2009-141
\end{flushright}
\vspace{15mm}
\begin{center}
{\LARGE Deformation Theory of Matrix Factorizations and F--Terms\\ \medskip in the Boundary Changing Sector} 
\end{center}
\vspace{10mm}
\begin{center}
{\large Johanna Knapp\footnote{email: \tt{knapp@mppmu.mpg.de}}}
\end{center}
\vspace{5mm}
\begin{center}
{\it Max--Planck--Institut f\"ur Physik \\   
F\"ohringer Ring 6\\
D--80805 Munich\\ 
Germany\\
}
\end{center}
\vspace{20mm}
\begin{abstract}
\noindent We extend the deformation theory algorithm of matrix factorizations to systems with more than one D--brane. The obstructions to the deformations are F--term equations which can be integrated to an effective superpotential. We demonstrate the power of this formalism for several examples of minimal models with two or three D--branes. These are realistic toy examples for systems with multiple D--branes in Calabi--Yau threefolds since both classes of models have obstructed open moduli.
\end{abstract}
\newpage
\setcounter{tocdepth}{1}
\tableofcontents
\setcounter{footnote}{0}
\section{Introduction}
Recently, there has been tremendous progress in mirror symmetry for open strings \cite{Walcher:2006rs,Morrison:2007bm,Krefl:2008sj,Knapp:2008uw,Jockers:2008pe,Grimm:2008dq,Alim:2009rf,Jockers:2009mn,Walcher:2009uj} on compact Calabi--Yau threefolds. By using various different techniques it was possible to calculate the effective superpotential $\mathcal{W}_{eff}$. Evaluating $\mathcal{W}_{eff}$ at pairs of critical points one obtains the domain--wall tension, which is the generating function of open Gromov--Witten invariants. The standard approach is to compute this quantity in the topological B--model and then make use of mirror symmetry to obtain the quantum corrected expression in the A--model. \\
So far, all the  mirror symmetry calculations for compact Calabi--Yau threefolds have been done for models with a single brane. What is of great phenomenological interest is to find methods to evaluate effective superpotential for models of intersecting branes and their mirrors. This is of course a very difficult problem and we will merely discuss small piece of this puzzle in this paper. When one looks at the mathematics literature, open string mirror symmetry is stated, very abstractly, as an equivalence of categories. However, it is this abstract formulation that makes it possible to treat systems with more than one brane and with boundary changing operators within the same framework as a single D--brane and its deformations. The problem with this approach is that one has to deal with a huge mathematical machinery which has a lot of structure but, in particular for the A--model, almost no technology which can be used to calculate concrete examples, in particular examples which are interesting for physics applications. In the topological B--model the situation is better than in the A--model since the derived category of coherent sheaves, which describes B--branes, is better understood than the Fukaya category, which describes the A--branes. If the example one is interested in has a Landau--Ginzburg phase, there is also a non--geometric description of B--branes in terms of matrix factorizations of the Landau--Ginzburg superpotential. One of the great advantages of matrix factorizations is that they give a very explicit realization of the category of D--branes. In particular one can discuss the boundary preserving and the boundary changing sector with the same methods. In many cases it is even not obvious whether a given matrix factorization describes a single D--brane or a bound state of several elementary branes. This is actually a big drawback of this formalism because one can classify matrix factorizations only for very simple models. However, in view of finding technologies which are equally powerful in the boundary preserving and the boundary changing sector the drawback becomes an advantage.\\
In \cite{Siqveland1,Knapp:2006rd,Knapp:2008tv} an algorithm for the deformation theory of matrix factorizations has been introduced. It computes the most general deformation of a matrix factorization (typically away from the Gepner point) modulo obstructions to these deformations. The obstructions are encoded in a set of equations which constrain the deformation parameters. These equations can be interpreted as F--term equations in the vicinity of the Gepner point \cite{Hori:2004ja}, and they can in fact be integrated to give the effective superpotential in the B--model. This algorithm was originally applied to topological minimal models. In \cite{Knapp:2008tv} it has been shown that the method can also be applied to Calabi--Yau threefolds. The reason that this works is that the open moduli of a Calabi--Yau threefold are generically obstructed\footnote{This property is specific for complex dimension $3$.}, which makes the Calabi--Yau behave very much like a minimal model, at least as far as deformation theory is concerned. Therefore minimal models serve as a very good playground for Calabi--Yau threefolds. The goal of this note is to extend the deformation theory algorithm to the boundary changing sector and show that it works for several minimal model examples. \\\\
The article is organized as follows: In section \ref{sec-massey} we extend the deformation theory algorithm to systems with more than one D--brane. We argue that the algorithm does not change at all if we just pack the matrix factorizations describing the branes of the model into a big matrix factorization. Since this can lead to very large matrices, we introduce an additional labelling into the formalism which allows us to break up the problem into the boundary preserving and boundary changing components, thus keeping the ranks of the matrices we deal with as small as in the case of a single D--brane. In section \ref{sec-a3} we discuss in some detail explicit examples for systems with two and three branes in the $A_4$ minimal model. In section \ref{sec-e6} we give an example for two branes in the $E_6$ minimal model. In section \ref{sec-cy} we outline a possible extension to Calabi--Yau threefolds. Finally, we present our conclusions in section \ref{sec-conc}.\\\\
{\bf Acknowledgments}: I would like to thank Emanuel Scheidegger for comments on the manuscript, and the organizers and participants of 4th Workshop on Symplectic Field Theory in Munich for a very interesting conference that inspired me to reconsider this project.
\section{The deformation theory algorithm for more than one brane}
\label{sec-massey}
We use the deformation/obstruction theory of matrix factorizations to compute F--terms equations near Gepner points. An algorithm for the systematic calculation of deformations of matrix factorizations has first been introduced in the mathematics literature in \cite{Siqveland1}. It goes under the name 'method of computing formal moduli' and has also been called Massey product algorithm, since it computes higher (Massey) products in the cohomology of a matrix factorization. For a single D--brane\footnote{In the non--geometric description the notion of what a single D--brane is is not so clear. Throughout the text we will work with matrix factorizations which can be mapped to certain boundary states in conformal field theory. These will be the constituent branes in the boundary changing models and we will refer to each of them as a 'single D--brane'.} this method has been demonstrated to work for minimal models \cite{Knapp:2006rd,Knapp:2007vc} and for Calabi--Yau hypersurfaces in weighted projective space \cite{Knapp:2008uw,Knapp:2008tv}. In this paper we extend the Massey product algorithm for systems with more than one D--brane. The extension turns out to be somewhat trivial, as we explain in the following.
\subsection{'Trivial' Extension}
One of the most intriguing features of viewing D--branes as objects in a category is that one can treat systems of several D--branes, bound states of branes and boundary changing open string states with the same techniques that are used for a single D--brane and boundary preserving open string states.
This feature is realized explicitly in matrix factorizations. Therefore it should not come as a surprise that the deformation theory of matrix factorizations describing a single D--brane works the same way for several branes. \\
A matrix factorization $Q_I$ associated to a brane with label $I$ is a matrix with polynomial entries satisfying
\begin{equation}
Q_I^2=W\cdot\mathbbm{1},
\end{equation}
where $W$ is the superpotential of the Landau--Ginzburg model. Matrix factorizations define a BRST operator on the boundary which determines the physical open string states. Given two matrix factorizations $Q_I$ and $Q_J$ of the same Landau--Ginzburg superpotential, the open string spectrum of defined by the cohomology of an operator $D^{(IJ)}$ which acts as follows on an open string state $\mathcal{S}$:
\begin{equation}
D^{(IJ)}\mathcal{S}\equiv Q_I\cdot\mathcal{S}-(-1)^{|\mathcal{S}|}\mathcal{S}\cdot Q_J
\end{equation}
Here $|\mathcal{S}|$ is the $\mathbbm{Z}_2$--degree of $\mathcal{S}$. Due to this grading one can divide up the states into $\mathbbm{Z}_2$--odd (fermionic) and $\mathbbm{Z}_2$--even (bosonic) open string states, depending on whether they anticommute or commute with $(Q_I,Q_J)$. The fermions $\Psi_i^{(IJ)}$ beginning on the brane represented by $Q_I$ and ending on the brane represented by $Q_J$ can be used to deform the matrix factorization. The bosonic open string states $\Phi_i^{(JI)}$ are the obstructions to these deformations. The index $i$ labels the various cohomology elements. For $I=J$, we are in the boundary preserving sector.\\
Given $N$ matrix factorizations $Q_A$  of ranks $r_A$ of a Landau--Ginzburg superpotential $W$ we can can associate a $\sum_Ar_A\times\sum_Ar_A$ matrix factorization to the system:
\begin{equation}
\label{nbranemf}
Q=\left(\begin{array}{ccccc}
Q_1&0&0&\cdots&0\\
0&Q_2&0&\cdots&0\\
0&0&Q_3&\cdots&0\\
\vdots&\vdots&\vdots&\ddots&\vdots\\
0&0&0&\cdots&Q_N
\end{array}\right)
\end{equation}
Obviously, $Q^2=W\cdot \mathbbm{1}_{\sum_{A=1}^Nr_A\times\sum_{A=1}^Nr_A}$. \\
The open string states $\Psi_i^{(IJ)}$ and $\Phi_i^{(JI)}$ determined by the $Q_A$ make up the arrows in a quiver diagram with $N$ nodes. To each fermionic open string state $\Psi_i^{(IJ)}$, we can associate a deformation parameter $u_i^{(IJ)}$ and deform (\ref{nbranemf}). This deformation will sit at the $(IJ)$th position in (\ref{nbranemf}). These open string states are  the linear deformations of the matrix factorization. Schematically, the deformed matrix factorization looks like this: 
\begin{equation}
\label{nbranemfdef}
Q_{def}^{lin}=\left(\begin{array}{ccccc}
Q_1+\sum_iu_i^{(11)}\Psi_i^{(11)}&\sum_iu_i^{(12)}\Psi_i^{(12)} &\sum_iu_i^{(13)}\Psi_i^{(13)} &\cdots&\sum_iu_i^{(1N)}\Psi_i^{(1N)}\\
\sum_iu_i^{(21)}\Psi_i^{(21)} &Q_2+\sum_iu_i^{(22)}\Psi_i^{(22)} &\sum_iu_i^{(23)}\Psi_i^{(23)} &\cdots&\sum_iu_i^{(2N)}\Psi_i^{(2N)} \\
\sum_iu_i^{(31)}\Psi_i^{(31)} &\sum_iu_i^{(32)}\Psi_i^{(32)} &Q_3+\sum_iu_i^{(33)}\Psi_i^{(33)} &\cdots&\sum_iu_i^{(3N)}\Psi_i^{(3N)} \\
\vdots&\vdots&\vdots&\ddots&\vdots\\
\sum_iu_i^{(N1)}\Psi_i^{(N1)} &\sum_iu_i^{(N2)}\Psi_i^{(N2)} &\sum_iu_i^{(N3)}\Psi_i^{(N3)} &\cdots&Q_N+\sum_iu_i^{(NN)}\Psi_i^{(NN)}
\end{array}\right)
\end{equation} 
Higher order deformations arise from imposing the matrix factorization condition modulo constraints on the deformation parameters. These define the critical locus of the effective superpotential. The matrix factorization (\ref{nbranemfdef}) has a very special block structure. However, this structure is not necessary for the definition of a matrix factorization and we may actually ignore it and apply the deformation theory algorithm just like for the case of a single brane. It turns out that this works without further modifications.\\
Even though we do not have to take into account the additional structure of the big matrix factorization, it may be wise to do so nonetheless. In this paper we only do explicit calculations for minimal models where the matrix factorizations are sufficiently small.  In \cite{Knapp:2008tv} it was demonstrated that the deformation theory algorithm also works very well for Calabi--Yau threefolds. A class of matrix factorizations which are convenient to work with are those which correspond to Recknagel--Schomerus boundary states in CFT. For a Calabi--Yau threefold which is a hypersurface in $(W)\mathbbm{CP}^4$ such branes are represented by $32\times32$ matrix factorizations. If one would like to consider, say, five Recknagel--Schomerus branes of the quintic which are an orbit of the diagonal $\mathbbm{Z}_5$ action, one would end up with a $160\times 160$ matrix. This is an inconveniently large matrix to work with. In the following section we will show how one can reduce the problem to calculations with $32\times 32$ matrices by introducing a new set of labels.
\subsection{New Labels}
We now show how we can make use of the special structure of the matrix factorization (\ref{nbranemfdef}) by introducing new labels into the Massey product algorithm and thus defining Massey products for subsectors of the big matrix factorization. In \cite{Jockers:2007ng} such an approach has been used to calculate threepoint functions on the torus. We make use of the fact that (\ref{nbranemfdef}) splits up into blocks which can be labeled by $(AB)$, marking the starting brane $Q_A$ and the end brane $Q_B$. Open string states starting on $Q_A$ and ending on $Q_B$ have zeroes everywhere but in the $(AB)$ sector of (\ref{nbranemfdef}). Their (Massey) products with other open string states are only non--zero if we multiply with a state of the sector $(*A)$ from the left of with an open string state of the sector $(B*)$ from the right, where the '$*$' can be any label. Products with non--matching labels are $0$. This property is automatically encoded in the big matrix (\ref{nbranemfdef}). If we want to work just with the blocks of (\ref{nbranemfdef}) we have to assign to every Massey product and to every higher order deformation a label $(AB)$ and impose the matching conditions. In terms of the quiver, these matching conditions mean that we must only consider connected paths in the quiver\footnote{Note that the paths have to be connected but need not be closed. The reason for this is that we are computing the F--terms and not the summands in the effective superpotential $\mathcal{W}_{eff}$. The F--terms are the derivatives of $\mathcal{W}_{eff}$ with respect to the brane moduli. These may have 'open ends' which are then closed by the integration. Then the picture matches with the interpretation of $\mathcal{W}_{eff}$ as the generating function of disk amplitudes.}.\\
In the following we will more or less copy the description of the Massey product algorithm as given in \cite{Knapp:2008tv} and systematically introduce the new labels.\\
Suppose the system of $N$ matrix factorizations has in total $M$ fermionic and $M$ bosonic open string states:
\begin{equation}
\Psi_k^{(AB)}\in H^{odd}(D^{(AB)})\qquad \Phi_k^{(AB)}\in H^{even}(D^{(AB)})\qquad \begin{array}{l}A,B\in \{1,\ldots, N\}\\ k=1,\ldots,M\end{array}
\end{equation}
where 
\begin{equation}
\sum_{A,B=1}^N\mathrm{dim}H^{even}(D^{(AB)})=\sum_{A,B=1}^N\mathrm{dim}H^{odd}(D^{(AB)})=M
\end{equation}
We associate deformation parameters $u_1,\ldots,u_M$ to the fermionic open string states. We would like to calculate the most general non--linear deformation of the large matrix factorization (\ref{nbranemf}), taking into account only deformations with $\mathbbm{Z}_2$--odd states. We make an ansatz for the higher order deformations. In the boundary preserving sector $A=B$, this looks as follows:
\begin{equation}
\label{eq-qdefans-bp}
Q_{A,def}=Q_A+\sum_{\vec{m}\in\bar{B}}\alpha^{(AA)}_{\vec{m}}u^{\vec{m}}
\end{equation}
In the boundary changing sector $A\neq B$ we have:
\begin{equation}
\label{eq-qdefans-bc}
\sum_{\vec{m}\in\bar{B}}\alpha^{(AB)}_{\vec{m}}u^{\vec{m}}
\end{equation}
Here, $\vec{m}$ is a multi index: $u^{\vec{m}}=u_1^{m_1}u_2^{m_2}\ldots u_M^{m_M}$, and we define $|\vec{m}|=\sum_{i=1}^M m_i$. $\bar{B}$ describes the allowed set of vectors $\vm$. Not all vectors $\vm$ are allowed due to relations coming from the F--terms. The matrices $\alpha^{(AB)}_{\vec{m}}$ will be determined recursively  in $|\vm|$. At the order $|\vec{m}|=1$ (linear deformations) they are defined to be the odd cohomology elements:
\begin{equation}
\label{alphadef}
\alpha^{(A_1B_1)}_{(1,0,\ldots,0)}=\Psi^{(A_1B_1)}_1\quad \alpha^{(A_2B_2)}_{(0,1,\ldots,0)}=\Psi^{(A_2B_2)}_2\quad\ldots\quad \alpha^{(A_MB_M)}_{(0,\ldots,0,1)}=\Psi^{(A_MB_M)}_{M} 
\end{equation}
The ans\"atze above are inserted at the $(A_iB_i)$th position in (\ref{nbranemf}).\\
We can then define enlarged matrices $\tilde{\alpha}_{\vm}^{(AB)}$  which are of size $\sum_{A=1}^Nr_A\times\sum_{A=1}^Nr_A$ and have zeros everywhere but in the $(AB)$--block of size $r_A\times r_B$. The general deformation of (\ref{nbranemf}) then looks like this:
\begin{equation}
\label{eq-qdefans}
Q_{def}=Q+\sum_{A,B=1}^N\sum_{\vec{m}\in\bar{B}}\tilde{\alpha}^{(AB)}_{\vec{m}}u^{\vec{m}}
\end{equation}
Imposing the matrix factorization condition on $Q_{def}$ we find:
\begin{eqnarray}
\label{eq-qdeffact1}
Q_{def}^2&\stackrel{!}{=}&W\cdot\mathbbm{1}+
\sum_{j=1}^M{\hat{f}}^{(A_jB_j)}_j(u)\tilde{\Phi}^{(A_jB_j)}_j\\
\label{eq-qdeffact2}
&\sim& Q^2+\sum_{A,B=1}^N\sum_{\vec{m}}[Q,\tilde{\alpha}^{(AB)}_{\vec{m}}]u^{\vec{m}}+\sum_{A,B,C,D=1}^N\sum_{\vec{m}_1+\vec{m}_2=\vec{m}}\underbrace{\tilde{\alpha}^{(AB)}_{\vec{m}_1}\cdot\tilde{\alpha}^{(CD)}_{\vec{m}_2}}_{y^{(ABCD)}(\vec{m})}u^{\vec{m}}
\end{eqnarray}
Here, the $\tilde{\Phi}^{(A_jB_J)}_j$ denote the enlarged bosonic open string states whose non--zero entries are in the $(A_jB_j)$ block.
As we discussed in the previous section, the additional labels introduced in the big matrix factorization are actually obsolete and can be removed: The labels $(A_jB_j)$ in (\ref{eq-qdeffact1}) are somewhat artificial since already the summation over $k$ accounts for all possible obstructions. Matrices $\tilde{\alpha}_{\vm}^{(AB)}$ with the same $\vm$--label but different $(AB)$ labels are collected into matrices which have entries in more than one $(AB)$--block. In the last summand of (\ref{eq-qdeffact2}) there are no constraints on the labels $A,B,C,D$. Any products of $\tilde{\alpha}$s which do not correspond to connected paths in the quiver will be $0$ automatically due to the structure of the matrices.\\
However, if we work with the small matrices $\alpha_{\vm}^{(AB)}$ the last term in (\ref{eq-qdeffact2}) will become:
\begin{equation}
y^{(AB)}(\vm)=\sum_C\sum_{\vec{m}_1+\vec{m}_2=\vec{m}}\alpha^{(AC)}_{\vec{m}_1}\cdot\alpha^{(CB)}_{\vec{m}_2}
\end{equation}
In this way we have thrown out all products which are trivially $0$ due to the block structure of the big matrix.\\
The products $y^{(AB)}(\vm)$ are called matric Massey products.  Obviously, the matrix factorization condition in (\ref{eq-qdeffact1}) only holds if we demand that ${\hat{f}}^{(A_kB_k)}_k(u)=0$. At the same time these relations determine the critical locus of the effective superpotential. Note that the $\hat{f}^{(A_kB_k)}_k(u)$ are equivalent to the obstructions $f^{(A_kB_k)}_k(u)$ which are built up by computing Massey products (see below).\\
The first Massey products $y^{(AB)}(\vm)$ appear at order $|\vm|=2$. We can calculate these products explicitly, since all the $\alpha^{(AB)}_{\vm}$ at order $|\vm|=1$ have been given in (\ref{alphadef}). $y^{(AB)}(\vm)$ can take the following values:
\begin{itemize}
\item $y^{(AB)}(\vm)$ is $D^{(AB)}$--exact. In this case we can find an $\alpha^{(AB)}_{\vm}$ with $|\vm|=2$ such that
\begin{equation}
Q_A\cdot\alpha^{(AB)}_{\vec{m}}+\alpha^{(AB)}_{\vec{m}}\cdot Q_B\equiv-\beta^{(AB)}_{\vm}=y^{(AB)}(\vec{m}).
\end{equation}
In this way we produce new $\alpha_{\vm}$'s and thus can calculate Massey products at higher order. 
\item $y^{(AB)}(\vm)\in H^{even}(D^{(AB)})$, i.e. $y^{(AB)}(\vm)=\sum_kc_k\:\Phi^{(AB)}_k$, where the $c_k$ is some numerical coefficients and the index $k$ goes over those labels in $\{1,\ldots,M\}$ which count bosonic open string states in the $(AB)$--sector. Clearly, this cannot be cancelled by a term $Q_A\cdot\alpha^{(AB)}_{\vec{m}}+\alpha^{(AB)}_{\vec{m}}\cdot Q_B $ since the $\Phi^{(AB)}_k$ are by definition not exact with respect to $D^{(AB)}$. Thus, we have encountered an obstruction. The obstructions are encoded in the polynomials $f^{(AB)}_k(u)$ associated to $\Phi^{(AB)}_k$ in the following way:
\begin{equation}
f^{(AB)}_k=c_k u^{\vm}
\end{equation}
\item There is also the possibility that  $y^{(AB)}(\vm)$ looks like an obstruction, i.e. it is not $D^{(AB)}$--exact, but cannot be associated to any bosonic open string state. Such terms do not spoil the algorithm and can be ignored during the calculation. In the end one can sum them up and gets constraints that look like F--terms. However, deformation theory tells us that there should only be as many F--terms as bosonic open string states. Therefore these extra relations must not contain new information but should be related to the regular F--term equations. Actually, one should even expect such contributions. The deformation theory yields a power series ring of deformations $\mathbbm{C}[[u]]/(f_1^{(A_1B_1)},\ldots,f_M^{(A_MB_M)})$ \cite{Siqveland1}. As long as the extra contributions are in the ideal generated by the F--terms nothing forbids their presence.\\
Such additional terms have already been observed in the boundary preserving sector in several examples of branes in Calabi--Yau threefolds \cite{Knapp:2008tv}. We will discuss their structure and properties in the example of the $A_4$ minimal model (cf. section \ref{sec-a3}). In all the examples we have computed, these additional equations are consistent with the F--terms.
\end{itemize}
At higher orders in $\vm$ we have to take into account the relations $f_j^{(A_jB_j)}$. This leads to more general definitions of Massey products and deformations. 
For a vector $\vn\in B'_{i+1}$, $i>0$ ($B'_{i+1}$ to be defined momentarily) the Massey product $y^{(AB)}(\vn)$ is given by:
\begin{equation}
\label{eq-massey}
y^{(AB)}(\vn)=\sum_C\sum_{|\vm|\leq i+1}\sum_{\stackrel{\vm_1+\vm_2=\vm}{\vm_i\in \bar{B}_{i}}}\beta'_{\vm,\vn}\alpha_{\vm_1}^{(AC)}\cdot\alpha^{(CB)}_{\vm_2}
\end{equation}
The coefficients $\beta'_{\vm,\vn}$ can be determined from the unique relation
\begin{equation}
\label{eq-masseycoeff1}
u^{\vn}=\sum_{\vm\in\bar{B}'_{i+1}}\beta'_{\vn,\vm}u^{\vm}+\sum_{j=1}^M\beta'_{\vn,j}f_j^{(A_jB_j)}
\end{equation}
for each $\vn\in\mathbbm{N}^M$ with $|\vn|\leq i+1$. If the Massey product is $y^{(AB)}(\vn)=\sum_kc_k\:\Phi^{(AB)}_k$ then we get a contribution to the polynomials $f^{(AB)}_k(u)$:
\begin{equation}
\label{eq-masseyobs}
f^{i+1,(AB)}_k=f_k^{i,(AB)}+\sum_{\vn\in B'_{i+1}}c_k\: u^{\vn},
\end{equation}
where the additional superscript gives the order in $u$.\\
The $\alpha^{(AB)}_{\vm}$ are defined as follows. For each vector $\vm$ in a basis $B_{i+1}$ we can find a matrix $\alpha^{(AB)}_{\vm}$  such that:
\begin{equation}
\label{eq-masseydef}
Q_A\cdot \alpha^{(AB)}_{\vm}+\alpha^{(AB)}_{\vm}\cdot Q_B=-\beta^{(AB)}_{\vm}=-\sum_{l=0}^{(i+1)-2}\sum_{\vn\in B'_{2+l}}\beta_{\vn,\vm}y^{(AB)}(\vn),
\end{equation}
where the coefficients $\beta_{\vn,\vm}$ are given by the unique relation
\begin{equation}
\label{eq-masseycoeff2}
u^{\vn}=\sum_{\vm\in\bar{B}_{i+1}}\beta_{\vn,\vm}u^{\vm}.
\end{equation}
The various bases $B$, $\bar{B}$, $B'$, $\bar{B}'$ are defined recursively. One starts by setting $\bar{B}_1=\{\vn\in\mathbbm{N}^{M}|\:|\vn|\leq1\}$ and $B_1=\{\vn\in\mathbbm{N}^M|\:|\vn|=1\}$. For $i\geq 1$, $B'_{i+1}$ is then defined as a basis for $m^{i+1}/(m^{i+2}+m^{i+1}\cap m(f^{i,(A_1B_1)}_1,\ldots,f^{i,(A_MB_M)}_M))$, where $m=(u_1,\ldots,u_M)$ defines the maximal ideal. In most cases, the elements $\{u^{\vn}\}_{\vn\in B'_{i+1}}$ can be chosen such that $u^{\vn}=u_k\cdot u^{\vm}$ for some $\vm\in \bar{B}_i$ and $u_k\in\{u_1,\ldots ,u_M\}$. One defines $\bar{B}'_{i+1}=\bar{B}_i\cup B'_{i+1}$. Finally, $B_{i+1}$ is a basis for $(m^{i+1}+(f_1^{i,(A_1B_1)},\ldots,f_M^{i,(A_MB_M)}))/(m^{i+2}+(f_1^{i+1,(A_1B_1)},\ldots,f_M^{i+1,(A_MB_M)}))$ such that $B_{i+1}\subseteq B'_{i+1}$. We set $\bar{B}_{i+1}=\bar{B}_i\cup B_{i+1}$. Note that the choice of bases does not have to be unique. \\
For minimal models one expects that the algorithm terminates at a certain order. For Calabi--Yau threefolds this need not happen but surprisingly does so in many examples \cite{Knapp:2008tv}.\\ 
In the following sections we will show how this algorithm works for some minimal model examples.
\subsection{Bulk deformations}
In \cite{Knapp:2008tv} we discussed how one can systematically add bulk deformations into the Massey product algorithm. If a bulk deformation can also be interpreted as a bosonic open string state, the associated bulk modulus contributes to the respective F--term. Otherwise the bulk deformation is $Q$--exact on the boundary. In this case one has to extend the vector $\vm$ by a further index associated to the bulk modulus and introduce a deformation $\alpha$ such that $\{Q,\alpha\}$ yields the bulk modulus. At higher orders the deformation theory algorithm works as usual with the extended index set which then also incorporates bulk deformations. \\
If we have more than one brane it may happen that the bulk deformation is $D^{(AB)}$--exact for one pair of branes but an obstruction for another combination. In this case we have to introduce a new index in $\vm$ for every bulk deformation which is $D^{(AB)}$--exact for at least one pair of matrix factorizations and a new deformation $\alpha_{\vm}^{(AB)}$ for every set of branes where the bulk deformation is $D^{(AB)}$--exact. Note that this holds in particular in the boundary preserving sector where $A=B$. However, if we consider a system where two or more constituent branes are equal, we also should expect bulk deformations in the boundary changing sector. After these modifications the algorithm proceeds as for the case without bulk deformations. In section \ref{sec-a3bulk} we compute an example for the $A_4$ minimal model with two branes and bulk deformations.
\section{The $A_4$ minimal model}
\label{sec-a3}
\subsection{Setup}
\begin{figure}
\begin{center}
\includegraphics{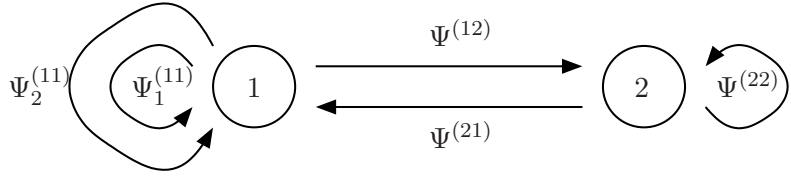}
\caption{The quiver diagram for the $A_4$ model with two branes.}\label{a3quiver-fig}
\end{center}
\end{figure}

Minimal models of type $A_{k+1}$ can be described in terms of a Landau--Ginzburg model with superpotential $W=x^{k+2}$. In this section we demonstrate how the extended Massey product algorithm works for two branes in the minimal model of type $A_4$. The Landau--Ginzburg superpotential is:
\begin{equation}
W=x^5
\end{equation}
The two matrix factorizations we are going to work with are:
\begin{equation}
\label{a4mf}
Q_1=\mat{0}{x^2}{x^3}{0}\qquad Q_2=\mat{0}{x}{x^4}{0}
\end{equation}
The system of the two branes is characterized by a $4\times 4$ matrix factorization which is the direct sum of the above two:
\begin{equation}
\label{qbig}
Q=\mat{Q_1}{0}{0}{Q_2}
\end{equation}
There are altogether five fermionic open string states which can be used to deform the matrix factorization. We label them with $\Psi^{(AB)}_i$, where $A$ and $B$ mark the start and endpoint of the open string state and $i$ is a label for the different open string states in the sector $(AB)$. In the boundary preserving sector of $Q_1$ there are two fermions with R--charges $\frac{1}{5}$ and $\frac{3}{5}$, respectively:
\begin{equation}
\Psi^{(11)}_1=\mat{0}{1}{-x}{0}\qquad \Psi^{(11)}_2=\mat{0}{x}{-x^2}{0}
\end{equation}
The boundary changing sector consists of two fermions of charge $\frac{1}{5}$:
\begin{equation}
\Psi^{(12)}=\mat{0}{1}{-x^2}{0}\qquad \Psi^{(21)}=\mat{0}{1}{-x^2}{0}
\end{equation}
Finally, there is one charge $\frac{3}{5}$ fermion in the boundary preserving sector of $Q_2$:
\begin{equation}
\Psi^{(22)}=\mat{0}{1}{-x^3}{0}
\end{equation}
Assigning a deformation parameter to each fermion, the linear deformation of $Q$ looks like this:
\begin{equation}
Q_{def}^{lin}=\mat{Q_1+u_2\Psi^{(11)}_1+u_1\Psi^{(11)}_2}{\tilde{u}_{\frac{3}{2}}\Psi^{(12)}}{\tilde{w}_{\frac{3}{2}}\Psi^{(21)}}{Q_2+w_1\Psi^{(22)}}
\end{equation}
Here, we have labeled the deformation parameters according to their weights. $(Q_{def}^{lin})^2$ yields the Massey products at second order in deformation theory. The obstructions to the deformations are encoded in the bosonic open string states, which we label by $\Phi_k^{(AB)}$. The two bosons in the boundary preserving sector of $Q_1$ have R--charges $0$ and $\frac{3}{5}$:
\begin{equation}
\Phi_1^{(11)}=\mat{1}{0}{0}{1}\qquad \Phi_2^{(11)}\mat{x}{0}{0}{x}
\end{equation}
The boundary changing bosons have charge $\frac{2}{5}$:
\begin{equation}
\Phi^{(12)}=\mat{x}{0}{0}{1}\qquad \Phi^{(21)}\mat{1}{0}{0}{x}
\end{equation}
Furthermore there is one charge $0$ boson in the boundary preserving sector of $Q_2$:
\begin{equation}
\Phi^{(22)}=\mat{1}{0}{0}{1}
\end{equation}
By Serre duality, the fermions pair up with the bosons such that the sum of the R--charges is $\frac{3}{5}$. Note that, in the boundary preserving sector, the Serre dual of a fermion $\Psi^{(AB)}$ will be a boson $\Phi^{(BA)}$. The open string spectrum can be encoded in the quiver diagram given in figure \ref{a3quiver-fig}.
In this figure we have only drawn the fermionic spectrum. The bosons are obtained by reversing all arrows.
\subsection{Deformation Theory}
We now have all the relevant ingredients for the deformation theory calculation. Besides the vector $\vec{m}$ which labels the order in the deformation parameters we also introduce a label $(*)^{(AB)}$ which labels the boundary preserving/changing sector, or equivalently, the position in the block matrix (\ref{qbig}). The deformations of the matrix factorizations are thus denoted by $\alpha^{(AB)}_{\vec{m}}$. The linear deformations are:
\begin{eqnarray}
\label{a3lin}
\alpha_{(1,0,0,0,0)}^{(11)}&=&\Psi_1^{(11)}\nonumber\\
\alpha_{(0,1,0,0,0)}^{(11)}&=&\Psi_2^{(11)}\nonumber\\
\alpha_{(0,0,1,0,0)}^{(12)}&=&\Psi^{(12)}\nonumber\\
\alpha_{(0,0,0,1,0)}^{(21)}&=&\Psi^{(21)}\nonumber\\
\alpha_{(0,0,0,0,1)}^{(22)}&=&\Psi^{(22)}
\end{eqnarray} 
Now we can compute the Massey products $y_{\vec{m}}^{(AB)}$ at order two in deformation theory, taking into account the new labels. At this order only one product hits an obstruction:
\begin{eqnarray}
y^{(11)}_{(2,0,0,0,0)}=\Psi_1^{(11)}\cdot \Psi_1^{(11)}=-\Phi_2^{(11)}
\end{eqnarray}
The remaining Massey products are $Q$--exact and therefore lead to new deformations. We group them according to their brane labels:
\begin{eqnarray}
y^{(11)}_{(1,1,0,0,0)}&=&\Psi_1^{(11)}\cdot \Psi_2^{(11)}+\Psi_2^{(11)}\cdot \Psi_1^{(11)}=-2x^2\cdot\mathbbm{1}\nonumber\\
y^{(11)}_{(0,2,0,0,0)}&=&\Psi_2^{(11)}\cdot \Psi_2^{(11)}=-x^3\cdot\mathbbm{1}\nonumber\\
y^{(11)}_{(0,0,1,1,0)}&=&\Psi^{(12)}\cdot \Psi^{(21)}=-x^2\cdot\mathbbm{1}
\end{eqnarray}
\begin{eqnarray}
y^{(12)}_{(1,0,1,0,0)}=\Psi_1^{(11)}\cdot \Psi^{(12)}=\mat{-x^2}{0}{0}{-x}\nonumber\\
y^{(12)}_{(0,1,1,0,0)}=\Psi_2^{(11)}\cdot \Psi^{(12)}=\mat{-x^3}{0}{0}{-x^2}\nonumber\\
y^{(12)}_{(0,0,1,0,1)}=\Psi^{(12)}\cdot \Psi^{(22)}=\mat{-x^3}{0}{0}{-x^2}
\end{eqnarray}
\begin{eqnarray}
y^{(21)}_{(1,0,0,1,0)}=\Psi^{(21)}\cdot\Psi_1^{(11)}=\mat{-x}{0}{0}{-x^2}\nonumber\\
y^{(21)}_{(0,1,0,1,0)}=\Psi^{(21)}\cdot\Psi_1^{(11)}=\mat{-x^2}{0}{0}{-x^3}\nonumber\\
y^{(21)}_{(0,0,0,1,1)}=\Psi^{(22)}\cdot\Psi_1^{(21)}=\mat{-x^2}{0}{0}{-x^3}
\end{eqnarray}
\begin{eqnarray}
y^{(22)}_{(0,0,1,1,0)}=\Psi^{(21)}\cdot\Psi^{(12)}=-x^2\cdot\mathbbm{1}\nonumber\\
y^{(22)}_{(0,0,0,0,2)}=\Psi^{(22)}\cdot\Psi^{(22)}=-x^3\cdot\mathbbm{1}
\end{eqnarray}
The $Q$--exact states are cancelled by the following deformations at order $2$:
\begin{eqnarray}
\label{a3order2}
\alpha^{(11)}_{(1,1,0,0,0)}&=&\mat{0}{0}{2}{0}\nonumber\\
\alpha^{(11)}_{(0,2,0,0,0)}&=&\mat{0}{a}{(1-a)x}{0}\nonumber\\
\alpha^{(11)}_{(0,0,1,1,0)}=\alpha^{(12)}_{(1,0,1,0,0)}=\alpha^{(21)}_{(1,0,0,1,0)}&=&\mat{0}{0}{1}{0}\nonumber\\
\alpha^{(12)}_{(0,1,1,0,0)}=\alpha^{(12)}_{(0,0,1,0,1)}=\alpha^{(21)}_{(0,1,0,1,0)}=\alpha^{(21)}_{(0,0,0,1,1)}=\alpha^{(22)}_{(0,0,1,1,0)}&=&\mat{0}{0}{x}{0}\nonumber\\
\alpha^{(22)}_{(0,0,0,0,2)}&=&\mat{0}{0}{x^2}{0}
\end{eqnarray}
Here we introduced a variable $a$ which parametrizes the choices is we can make for the deformations. At this order, the F--term equations looks as follows:
\begin{eqnarray}
f^{2,(11)}_1:&\quad&0\nonumber\\
f^{2,(11)}_2:&\quad&-u_2^2\nonumber\\
f^{2,(12)}:&\quad&0\nonumber\\
f^{2,(21)}:&\quad&0\nonumber\\
f^{2,(22)}:&\quad&0
\end{eqnarray}
Since the F--terms only account for a reduction of the number of $\vm$ with $|\vm|=3$ in the basis $B'_3$ but do not generate any extra terms in the Massey products, the calculation of the products at order $3$ is straight forward, and we refrain from writing down the components of the products. At order $3$, eleven Massey products lead to contributions to the F--terms:
\begin{eqnarray}
y^{(11)}_{(1,0,1,1,0)}&=&2\Phi_1^{(11)}\nonumber\\
y^{(11)}_{(1,2,0,0,0)}&=&(3-2a)\Phi^{(11)}_2\nonumber\\
y^{(11)}_{(0,1,1,1,0)}=y^{(11)}_{(0,0,1,1,1)}&=&2\Phi^{(11)}_2\nonumber\\
y^{(12)}_{(1,1,1,0,0)}=2y^{(12)}_{(1,0,1,0,1)}=2y^{(12)}_{(0,0,2,1,0)}&=&2\Phi^{(12)}\nonumber\\
y^{(21)}_{(1,1,0,1,0)}=2y^{(21)}_{(1,0,0,1,1)}=2y^{(21)}_{(0,0,1,2,0)}&=&2\Phi^{(21)}\nonumber\\
y^{(22)}_{(1,0,1,1,0)}&=&\Phi^{(22)}
\end{eqnarray}
Furthermore, ten products lead to $Q$--exact expressions:
\begin{eqnarray}
y^{(11)}_{(0,3,0,0,0)}&=&(1-2a)x^2\cdot\mathbbm{1}\nonumber\\
y^{(12)}_{(0,1,1,0,1)}=y^{(12)}_{(0,0,1,0,2)}=(1-2a)^{-1}y^{(12)}_{(0,2,1,0,0)}&=&\mat{x^2}{0}{0}{x}\nonumber\\
y^{(21)}_{(0,1,0,1,1)}=y^{(21)}_{(0,0,0,1,2)}=(1-a)^{-1}y^{(21)}_{(0,2,0,1,0)}&=&\mat{x}{0}{0}{x^2}\nonumber\\
y^{(22)}_{(0,0,1,1,1)}=2y^{(22)}_{(0,1,1,1,0)}&=&2x\cdot\mathbbm{1}\nonumber\\
y^{(22)}_{(0,0,0,0,3)}&=&x^2\cdot\mathbbm{1}
\end{eqnarray}
This leads to ten deformations at order $3$:
\begin{eqnarray}
\label{a3order3}
\alpha^{(11)}_{(0,3,0,0,0)}&=&-(1-2a)\mat{0}{0}{1}{0}\nonumber\\
\alpha^{(12)}_{(0,1,1,0,1)}=\alpha^{(12)}_{(0,0,1,0,2)}=(1-a)^{-1}\alpha^{(12)}_{(0,2,1,0,0)}&=&-\mat{0}{0}{1}{0}\nonumber\\
\alpha^{(21)}_{(0,1,0,1,1)}=\alpha^{(21)}_{(0,0,0,1,2)}=(1-a)^{-1}\alpha^{(21)}_{(0,2,0,1,0)}&=&-\mat{0}{0}{1}{0}\nonumber\\
\alpha^{(22)}_{(0,0,1,1,1)}=2\alpha^{(22)}_{(0,1,1,1,0)}&=&-2\mat{0}{0}{1}{0}\nonumber\\
\alpha^{(22)}_{(0,0,0,0,3)}&=&-x\mat{0}{0}{1}{0}
\end{eqnarray}
The F--term equations acquire several new terms:
\begin{eqnarray}
f^{3,(11)}_1:&\quad&2u_2\tilde{u}_{\frac{3}{2}}\tilde{w}_{\frac{3}{2}}\nonumber\\
f^{3,(11)}_2:&\quad&-u_2^2+(3-2a)u_2u_1^2+2u_1\tilde{u}_{\frac{3}{2}}\tilde{w}_{\frac{3}{2}}+\tilde{u}_{\frac{3}{2}}\tilde{w}_{\frac{3}{2}}w_1\nonumber\\
f^{3,(12)}:&\quad&2u_2u_1\tilde{u}_{\frac{3}{2}}+u_2\tilde{u}_{\frac{3}{2}}w_1+\tilde{u}^2_{\frac{3}{2}}\tilde{w}_{\frac{3}{2}}\nonumber\\
f^{3,(21)}:&\quad&2u_2u_1\tilde{w}_{\frac{3}{2}}+u_2\tilde{w}_{\frac{3}{2}}w_1+\tilde{u}_{\frac{3}{2}}\tilde{w}^2_{\frac{3}{2}} \nonumber\\
f^{3,(22)}:&\quad&u_2\tilde{u}_{\frac{3}{2}}\tilde{w}_{\frac{3}{2}}
\end{eqnarray}
At order $4$ the calculation becomes tedious due to the large number of $\alpha$'s and the fact that we get relations between the vectors $\vec{m}$ from the F--terms. We only explicitly write out those products where such relations contribute.\\
There are sixteen products which contribute to the F--term equations:
\begin{eqnarray}
y^{(11)}_{(1,3,0,0,0)}&=&\{\alpha^{(11)}_{(1,2,0,0,0)},\alpha^{(11)}_{(0,2,0,0,0)}\}+\{\alpha^{(11)}_{(0,3,0,0,0)},\Psi^{(11)}_1\}+(3-2a)\{\alpha^{(11)}_{(1,1,0,0,0)},\Psi^{(11)}_{1}\}=5\Phi^{(11)}_1\nonumber \\
y^{(11)}_{(0,1,1,1,1)}&=&\alpha^{(12)}_{(0,1,1,0,0)}\cdot\alpha^{(21)}_{(0,0,0,1,1)}+\alpha^{(12)}_{(0,0,1,0,1)}\cdot\alpha^{(21)}_{(0,1,0,1,0)}+\alpha^{(12)}_{(0,1,1,0,1)}\cdot\Psi^{(21)}+\nonumber\\
&&+\Psi^{(12)}\cdot\alpha^{(21)}_{(0,1,0,1,1)}+\{\alpha^{(11)}_{(1,1,0,0,0)},\Psi^{(11)}_1\}=\Phi_1^{(11)}\nonumber\\
y^{(11)}_{(0,2,1,1,0)}&=&\{\alpha^{(11)}_{(0,2,0,0,0)},\alpha^{(11)}_{(0,0,1,1,0)}\}+\alpha^{(12)}_{(0,1,1,0,0)}\cdot\alpha^{(21)}_{(0,1,0,1,0)}+\alpha^{(12)}_{(0,2,1,0,0)}\cdot\Psi^{(21)}\nonumber\\
&&+2\{\alpha^{(11)}_{(1,1,0,0,0)},\Psi_1^{(11)}\}=(3+2a)\Phi_1^{(11)}
\end{eqnarray} 
\begin{eqnarray}
y^{(11)}_{(0,4,0,0,0)}&=&-(1-3a+a^2)\Phi^{(11)}_2\nonumber\\
y^{(11)}_{(0,0,1,1,2)}&=&-\Phi_1^{(11)}\nonumber\\
y^{(12)}_{(0,1,1,0,2)}=y^{(12)}_{(0,0,1,0,3)}=(1-2a)^{-1}y^{(12)}_{(0,3,1,0,0)}=(1-a)^{-1}y^{(12)}_{(0,2,1,0,1)}&=&-\Phi^{(12)}\nonumber\\
y^{(21)}_{(0,1,0,1,2)}=y^{(21)}_{(0,0,0,1,3)}=(1-2a)^{-1}y^{(21)}_{(0,3,0,1,0)}=(1-a)^{-1}y^{(21)}_{(0,2,0,1,1)}&=&-\Phi^{(21)}\nonumber\\
(1-a)y^{(22)}_{(0,2,1,1,0)}=\frac{1}{2}y^{(22)}_{(0,1,1,1,1)}=\frac{1}{3}y^{(22)}_{(0,0,1,1,2)}&=&-\Phi^{(22)}
\end{eqnarray}
There is only one Massey product which yields something $Q$--exact:
\begin{equation}
y^{(22)}_{(0,0,0,0,4)}=-x\cdot\mathbbm{1}
\end{equation}
This in cancelled by the following deformation:
\begin{equation}
\label{a3order4}
\alpha^{(22)}_{(0,0,0,0,4)}=\mat{0}{0}{1}{0}
\end{equation}
The F--terms have grown significantly at order $4$:
\begin{eqnarray}
f^{4,(11)}_1:&\quad&2u_2\tilde{u}_{\frac{3}{2}}\tilde{w}_{\frac{3}{2}}+5u_2u_1^3+(3+2a)u_1^2\tilde{u}_{\frac{3}{2}}\tilde{w}_{\frac{3}{2}}+u_1\tilde{u}_{\frac{3}{2}}\tilde{w}_{\frac{3}{2}}w_1-\tilde{u}_{\frac{3}{2}}\tilde{w}_{\frac{3}{2}}w_1^2\nonumber\\
f^{4,(11)}_2:&\quad&-u_2^2+(3-2a)u_2u_1^2+2u_1\tilde{u}_{\frac{3}{2}}\tilde{w}_{\frac{3}{2}}+\tilde{u}_{\frac{3}{2}}\tilde{w}_{\frac{3}{2}}w_1-(1-3a+a^2)u_1^4\nonumber\\
f^{4,(12)}:&\quad&2u_2u_1\tilde{u}_{\frac{3}{2}}+u_2\tilde{u}_{\frac{3}{2}}w_1+\tilde{u}^2_{\frac{3}{2}}\tilde{w}_{\frac{3}{2}}-(1-2a)u_1^3\tilde{u}_{\frac{3}{2}}-(1-a)u_1^2\tilde{u}_{\frac{3}{2}}w_1-u_1\tilde{u}_{\frac{3}{2}}w_1^2-\tilde{u}_{\frac{3}{2}}w_1^3\nonumber\\
f^{4,(21)}:&\quad&2u_2u_1\tilde{w}_{\frac{3}{2}}+u_2\tilde{w}_{\frac{3}{2}}w_1+\tilde{u}_{\frac{3}{2}}\tilde{w}^2_{\frac{3}{2}}-(1-2a)u_1^3\tilde{w}_{\frac{3}{2}}-(1-a)u_1^2\tilde{w}_{\frac{3}{2}}w_1-u_1\tilde{w}_{\frac{3}{2}}w_1^2-\tilde{w}_{\frac{3}{2}}w_1^3 \nonumber\\
f^{4,(22)}:&\quad&u_2\tilde{u}_{\frac{3}{2}}\tilde{w}_{\frac{3}{2}}-(1-a)u_1^2\tilde{u}_{\frac{3}{2}}\tilde{w}_{\frac{3}{2}}-2u_1\tilde{u}_{\frac{3}{2}}\tilde{w}_{\frac{3}{2}}w_1-3\tilde{u}_{\frac{3}{2}}\tilde{w}_{\frac{3}{2}}w_1^2
\end{eqnarray}
At order $4$ we also find terms which appear to be problematic at first sight since they are neither $Q$--exact, nor are they proportional to bosonic open sting states. These exceptional Massey products do not influence the algorithm at any way and can be ignored throughout the calculation. We will discuss their consistency in section \ref{sec-a3supo}.
\begin{eqnarray}
y^{(12)}_{(1,0,1,0,2)}&=&\alpha^{(12)}_{(1,0,1,0,0)}\cdot\alpha^{(22)}_{(0,0,0,0,2)}+\Psi^{(11)}_1\cdot\alpha^{(12)}_{(0,0,1,0,2)}-\frac{1}{2}\left(\alpha^{(11)}_{(1,1,0,0,0)}\cdot\alpha^{(12)}_{(0,0,1,0,1)}+\Psi^{(11)}_1\cdot\alpha^{(12)}_{(0,1,1,0,1)}\right)\nonumber\\
&=&\mat{-\frac{1}{2}}{0}{0}{0}\nonumber\\
y^{(21)}_{(1,0,0,1,2)}&=&\alpha^{(22)}_{(0,0,0,0,2)}\cdot\alpha^{(21)}_{(1,0,0,1,0)}+\alpha^{(21)}_{(0,0,0,1,2)}\cdot\Psi^{(11)}_1-\frac{1}{2}\left(\alpha^{(21)}_{(0,0,0,1,1)}\cdot\alpha^{(11)}_{(1,1,0,0,0)}+\alpha^{(21)}_{(0,1,0,1,1)}\cdot\Psi^{(11)}_1\right)\nonumber\\
&=&\mat{0}{0}{0}{-\frac{1}{2}}\nonumber\\
y^{(12)}_{(0,1,2,1,0)}&=&\alpha^{(12)}_{(0,1,1,0,0)}\cdot\alpha^{(22)}_{(0,0,1,1,0)}+\alpha^{(11)}_{(0,0,1,1,0)}\cdot\alpha^{(12)}_{(0,1,1,0,0)}+\Psi^{(12)}\cdot\alpha^{(22)}_{(0,1,1,1,0)}+2\Psi^{(11)}_1\cdot\alpha^{(12)}_{(1,0,1,0,0)}\nonumber\\
&=&\mat{1}{0}{0}{0}\nonumber\\
y^{(21)}_{(0,1,1,2,0)}&=&\alpha^{(21)}_{(0,1,0,1,0)}\cdot\alpha^{(11)}_{(0,0,1,1,0)}+\alpha^{(22)}_{(0,0,1,1,0)}\cdot\alpha^{(21)}_{(0,1,0,1,0)}+2\alpha^{(21)}_{(1,0,0,1,0)}\cdot\Psi^{(11)}_1\nonumber\\
&=&\mat{0}{0}{0}{1}\nonumber\\
y^{(12)}_{(0,0,2,1,1)}&=&\alpha^{(11)}_{(0,0,1,1,0)}\cdot\alpha^{(12)}_{(0,0,1,0,1)}+\alpha^{(12)}_{(0,0,1,0,1)}\cdot\alpha^{(22)}_{(0,0,1,1,0)}+\Psi^{(12)}\cdot\alpha^{(22)}_{(0,0,1,1,1)}+\Psi^{(11)}_1\cdot\alpha^{(12)}_{(1,0,1,0,0)}\nonumber\\
&&-\frac{1}{2}\left(\alpha^{(11)}_{(1,1,0,0,0)}\cdot\alpha^{(12)}_{(0,0,1,0,1)}+\Psi^{(11)}_1\cdot\alpha^{(12)}_{(0,1,1,0,1)}\right)=\mat{-\frac{1}{2}}{0}{0}{0}\nonumber\\
y^{(21)}_{(0,0,1,2,1)}&=&\alpha^{(22)}_{(0,0,1,1,0)}\cdot\alpha^{(21)}_{(0,0,0,1,1)}+\alpha^{(21)}_{(0,0,0,1,1)}\cdot\alpha^{(11)}_{(0,0,1,1,0)}+\alpha^{(22)}_{(0,0,1,1,1)}\cdot\Psi^{(21)}+\alpha^{(21)}_{(1,0,0,1,0)}\cdot\Psi^{(11)}_1\nonumber\\
&&-\frac{1}{2}\left(\alpha^{(21)}_{(0,0,0,1,1)}\cdot\alpha^{(11)}_{(1,1,0,0,0)}+\alpha^{(21)}_{(0,1,0,1,1)}\cdot\Psi^{(11)}_1\right)=\mat{0}{0}{0}{-\frac{1}{2}}\nonumber\\
y^{(12)}_{(1,2,1,0,0)}&=&\alpha^{(11)}_{(0,2,0,0,0)}\cdot\alpha^{(12)}_{(1,0,1,0,0)}+\alpha^{(11)}_{(1,1,0,0,0)}\cdot\alpha^{(12)}_{(0,1,1,0,0)}+\Psi^{(11)}_1\cdot\alpha^{(12)}_{(0,2,1,0,0)}\nonumber\\
&&+(3-2a)(\Psi^{(11)}_1\cdot\alpha^{(12)}_{(1,0,1,0,0)})=\mat{2}{0}{0}{0}\nonumber\\
y^{(21)}_{(1,2,0,1,0)}&=&\alpha^{(21)}_{(1,0,0,1,0)}\cdot\alpha^{(11)}_{(0,2,0,0,0)}+\alpha^{(21)}_{(0,1,0,1,0)}\cdot\alpha^{(11)}_{(1,1,0,0,0)}+\alpha^{(21)}_{(0,2,0,1,0)}\cdot\Psi^{(11)}_1\nonumber\\
&&+(3-2a)(\alpha^{(21)}_{(1,0,0,1,0)}\cdot\Psi^{(11)}_1)=\mat{0}{0}{0}{2}
\end{eqnarray}
At order $5$ in deformation theory, the majority of Massey products is $0$. There are only two which contribute to the obstructions:
\begin{eqnarray}
y^{(11)}_{(0,5,0,0,0)}&=&\{\alpha^{(11)}_{(0,2,0,0,0)},\alpha^{(11)}_{(0,3,0,0,0)}\}-(1-3a+a^2)\{\alpha^{(11)}_{(1,1,0,0,0)},\Psi^{(11)}_1\}=-(2-5a)\Phi^{(11)}_1\nonumber\\
y^{(22)}_{(0,0,0,0,5)}&=&\{\alpha^{(22)}_{(0,0,0,0,2)},\alpha^{(22)}_{(0,0,0,0,3)}\}+\{\alpha^{(22)}_{(0,0,0,0,4)},\Psi^{(22)}\}=\Phi^{(22)}
\end{eqnarray}
The F--terms at order $5$ are:
\begin{eqnarray}
f^{5,(11)}_1:&\quad&2u_2\tilde{u}_{\frac{3}{2}}\tilde{w}_{\frac{3}{2}}+5u_2u_1^3+(3+2a)u_1^2\tilde{u}_{\frac{3}{2}}\tilde{w}_{\frac{3}{2}}+u_1\tilde{u}_{\frac{3}{2}}\tilde{w}_{\frac{3}{2}}w_1-\tilde{u}_{\frac{3}{2}}\tilde{w}_{\frac{3}{2}}w_1^2-(2-5a)u_1^5\nonumber\\
f^{5,(11)}_2:&\quad&-u_2^2+(3-2a)u_2u_1^2+2u_1\tilde{u}_{\frac{3}{2}}\tilde{w}_{\frac{3}{2}}+\tilde{u}_{\frac{3}{2}}\tilde{w}_{\frac{3}{2}}w_1-(1-3a+a^2)u_1^4\nonumber\\
f^{5,(12)}:&\quad&2u_2u_1\tilde{u}_{\frac{3}{2}}+u_2\tilde{u}_{\frac{3}{2}}w_1+\tilde{u}^2_{\frac{3}{2}}\tilde{w}_{\frac{3}{2}}-(1-2a)u_1^3\tilde{u}_{\frac{3}{2}}-(1-a)u_1^2\tilde{u}_{\frac{3}{2}}w_1-u_1\tilde{u}_{\frac{3}{2}}w_1^2-\tilde{u}_{\frac{3}{2}}w_1^3\nonumber\\
f^{5,(21)}:&\quad&2u_2u_1\tilde{w}_{\frac{3}{2}}+u_2\tilde{w}_{\frac{3}{2}}w_1+\tilde{u}_{\frac{3}{2}}\tilde{w}^2_{\frac{3}{2}}-(1-2a)u_1^3\tilde{w}_{\frac{3}{2}}-(1-a)u_1^2\tilde{w}_{\frac{3}{2}}w_1-u_1\tilde{w}_{\frac{3}{2}}w_1^2-\tilde{w}_{\frac{3}{2}}w_1^3 \nonumber\\
f^{5,(22)}:&\quad&u_2\tilde{u}_{\frac{3}{2}}\tilde{w}_{\frac{3}{2}}-(1-a)u_1^2\tilde{u}_{\frac{3}{2}}\tilde{w}_{\frac{3}{2}}-2u_1\tilde{u}_{\frac{3}{2}}\tilde{w}_{\frac{3}{2}}w_1-3\tilde{u}_{\frac{3}{2}}\tilde{w}_{\frac{3}{2}}w_1^2+w_1^5
\end{eqnarray}
Again, there are some problematic expressions which are neither deformations nor true obstructions:
\begin{eqnarray}
&y^{(12)}_{(0,4,1,0,0)}=\mat{-1+2a}{0}{0}{0}\qquad &y^{(21)}_{(0,4,0,1,0,0)}=\mat{0}{0}{0}{-1+2a}\nonumber\\
&y^{(12)}_{(0,3,1,0,1)}=\mat{-\frac{1}{2}}{0}{0}{0}\qquad &y^{(21)}_{(0,3,0,1,0,1)}=\mat{0}{0}{0}{-\frac{1}{2}}\nonumber\\
&y^{(12)}_{(0,2,1,0,2)}=\mat{-\frac{1+a}{2}}{0}{0}{0}\qquad &y^{(21)}_{(0,2,0,1,0,2)}=\mat{0}{0}{0}{-\frac{1+a}{2}}\nonumber\\
&y^{(12)}_{(0,1,1,0,3)}=\mat{-\frac{1}{2}}{0}{0}{0}\qquad &y^{(21)}_{(0,1,0,1,0,3)}=\mat{0}{0}{0}{-\frac{1}{2}}\nonumber\\
&y^{(12)}_{(0,0,1,0,4)}=\mat{\frac{1}{2}}{0}{0}{0}\qquad &y^{(21)}_{(0,0,0,1,0,4)}=\mat{0}{0}{0}{\frac{1}{2}}
\end{eqnarray}
All the Massey products at orders $6$, $7$ and $8$ are $0$. From order $9$ on, there are no $\alpha$'s left we could multiply. Thus, the algorithm terminates. 
\subsection{Effective Superpotential}
\label{sec-a3supo}
The full F--term equations are:
\begin{eqnarray}
\label{a3fterm}
f^{(11)}_1:&\quad&2u_2\tilde{u}_{\frac{3}{2}}\tilde{w}_{\frac{3}{2}}+5u_2u_1^3+(3+2a)u_1^2\tilde{u}_{\frac{3}{2}}\tilde{w}_{\frac{3}{2}}+u_1\tilde{u}_{\frac{3}{2}}\tilde{w}_{\frac{3}{2}}w_1-\tilde{u}_{\frac{3}{2}}\tilde{w}_{\frac{3}{2}}w_1^2-(2-5a)u_1^5\nonumber\\
f^{(11)}_2:&\quad&-u_2^2+(3-2a)u_2u_1^2+2u_1\tilde{u}_{\frac{3}{2}}\tilde{w}_{\frac{3}{2}}+\tilde{u}_{\frac{3}{2}}\tilde{w}_{\frac{3}{2}}w_1-(1-3a+a^2)u_1^4\nonumber\\
f^{(12)}:&\quad&2u_2u_1\tilde{u}_{\frac{3}{2}}+u_2\tilde{u}_{\frac{3}{2}}w_1+\tilde{u}^2_{\frac{3}{2}}\tilde{w}_{\frac{3}{2}}-(1-2a)u_1^3\tilde{u}_{\frac{3}{2}}-(1-a)u_1^2\tilde{u}_{\frac{3}{2}}w_1-u_1\tilde{u}_{\frac{3}{2}}w_1^2-\tilde{u}_{\frac{3}{2}}w_1^3\nonumber\\
f^{(21)}:&\quad&2u_2u_1\tilde{w}_{\frac{3}{2}}+u_2\tilde{w}_{\frac{3}{2}}w_1+\tilde{u}_{\frac{3}{2}}\tilde{w}^2_{\frac{3}{2}}-(1-2a)u_1^3\tilde{w}_{\frac{3}{2}}-(1-a)u_1^2\tilde{w}_{\frac{3}{2}}w_1-u_1\tilde{w}_{\frac{3}{2}}w_1^2-\tilde{w}_{\frac{3}{2}}w_1^3 \nonumber\\
f^{(22)}:&\quad&u_2\tilde{u}_{\frac{3}{2}}\tilde{w}_{\frac{3}{2}}-(1-a)u_1^2\tilde{u}_{\frac{3}{2}}\tilde{w}_{\frac{3}{2}}-2u_1\tilde{u}_{\frac{3}{2}}\tilde{w}_{\frac{3}{2}}w_1-3\tilde{u}_{\frac{3}{2}}\tilde{w}_{\frac{3}{2}}w_1^2+w_1^5
\end{eqnarray}
The equations have weights $\{5,4,\frac{9}{2},\frac{9}{2},5\}$, respectively. One can show \cite{Knapp:2006rd,Knapp:2007vc} that the effective superpotential for this model has weight $6$. The integration variables have weights $\{1,2,\frac{3}{2},\frac{3}{2},1\}$. To integrate, we have to combine the F--term equations such that the weight of the combination plus the weight of the integration variable equals $6$. Up to an overall factor, the unknown parameters are determined by the condition that the second order derivatives of the integral have to match pairwise. This yields the following result:
\begin{eqnarray}
\mathcal{W}_{eff}&=&\frac{1-6a+9a^2-2a^3}{6}u_1^6+(-1+3a-a^2)u_1^4u_2+\frac{3-2a}{2}u_1^2u_2^2-\frac{1}{3}u_2^2\nonumber\\
&&-(1-2a)u_1^3\tilde{u}_{\frac{3}{2}}\tilde{w}_{\frac{3}{2}}+2u_1u_2\tilde{u}_{\frac{3}{2}}\tilde{w}_{\frac{3}{2}}+\frac{1}{2}\tilde{u}_{\frac{3}{2}}^2\tilde{w}_{\frac{3}{2}}^2-(1-a)u_1^2\tilde{u}_{\frac{3}{2}}\tilde{w}_{\frac{3}{2}}w_1+u_2\tilde{u}_{\frac{3}{2}}\tilde{w}_{\frac{3}{2}}w_1\nonumber\\
&&-u_1\tilde{u}_{\frac{3}{2}}\tilde{w}_{\frac{3}{2}}w_1^2-\tilde{u}_{\frac{3}{2}}\tilde{w}_{\frac{3}{2}}w_1^3+\frac{1}{6}w_1^6
\end{eqnarray}
For $a=0$ and $\{u_1\rightarrow-u_1,u_2\rightarrow-u_2,w_1\rightarrow-w_1\}$ this agrees with the result in \cite{Herbst:2004jp}, which was found by solving consistency conditions for disk amplitudes. Solutions of the F--term equations for two branes of minimal models of type $A$ have been discussed in \cite{Herbst:2004zm}. They contain the information about which bound states can be formed by topological tachyon condensation, depending on which solution of (\ref{a3fterm}) is chosen.
\subsubsection{Extra terms and consistency}
Collecting all the expressions for the deformations $\alpha^{(AB)}_{\vm}$ from (\ref{a3lin}), (\ref{a3order2}), (\ref{a3order3}) and (\ref{a3order4}) we can explicitly compute the deformed matrix factorization of the two--brane system. It has the following structure:
\begin{equation}
Q_{def}=\mat{Q_1+\sum_{\vm}\alpha^{(11)}_{\vm}u^{\vm}}{\sum_{\vm}\alpha^{(12)}_{\vm}u^{\vm}}{\sum_{\vm}\alpha^{(21)}_{\vm}u^{\vm}}{Q_2+\sum_{\vm}\alpha^{(22)}_{\vm}u^{\vm}}
\end{equation}
Squaring this matrix, we get the following expression:
\begin{equation}
Q_{def}^2=\mat{W\cdot\mathbbm{1}_{2\times 2}+f_2^{(11)}\Phi_2^{(11)}+(f_1^{(11)}-2u_1f_2^{(11)})\Phi_1^{(11)}}{f^{(12)}\Phi^{(12)}}{f^{(21)}\Phi^{(21)}}{W\cdot\mathbbm{1}_{2\times 2}+f^{(22)}\Phi^{(22)}}+X,
\end{equation}
where 
\begin{equation}
X=\left(\begin{array}{rr}
\mat{0}{0}{0}{0}&\tilde{f}^{(12)}\mat{1}{0}{0}{0}\nonumber\\
\tilde{f}^{(21)}\mat{0}{0}{0}{1}&\mat{0}{0}{0}{0}
\end{array}\right),
\end{equation}
with new obstruction--like terms:
\begin{eqnarray}
\tilde{f}^{(12)}&=&u_2^2\tilde{u}_{\frac{3}{2}}-(1-2a)u_1^2u_2\tilde{u}_{\frac{3}{2}}-a(1-a)u_1^4\tilde{u}_{\frac{3}{2}}-u_1\tilde{u}^2_{\frac{3}{2}}\tilde{w}_{\frac{3}{2}}-u_1u_2\tilde{u}_{\frac{3}{2}}w_1-au_1^3\tilde{u}_{\frac{3}{2}}\nonumber\\
&&-2\tilde{u}^2_{\frac{3}{2}}\tilde{w}_{\frac{3}{2}}w_1-u_2\tilde{u}_{\frac{3}{2}}w_1^2-au_1^2\tilde{u}_{\frac{3}{2}}w_1^2+\tilde{u}_{\frac{3}{2}}w_1^4\nonumber\\
\tilde{f}^{(21)}&=&u_2^2\tilde{w}_{\frac{3}{2}}-(1-2a)u_1^2u_2\tilde{w}_{\frac{3}{2}}-a(1-a)u_1^4\tilde{w}_{\frac{3}{2}}-u_1\tilde{u}_{\frac{3}{2}}\tilde{w}^2_{\frac{3}{2}}-u_1u_2\tilde{w}_{\frac{3}{2}}w_1-au_1^3\tilde{w}_{\frac{3}{2}}\nonumber\\
&&-2\tilde{u}_{\frac{3}{2}}\tilde{w}^2_{\frac{3}{2}}w_1-u_2\tilde{w}_{\frac{3}{2}}w_1^2-au_1^2\tilde{w}_{\frac{3}{2}}w_1^2+\tilde{w}_{\frac{3}{2}}w_1^4
\end{eqnarray}
At first sight these expressions look like new F--terms. This would be in contradiction to deformation theory which tells us that there are as many obstructions as there are bosonic open string states. So, these equations had better be consistent with the F--terms equations. Indeed, a small calculation shows that we have:
\begin{eqnarray}
\tilde{f}^{(12)}&=&-\tilde{u}_{\frac{3}{2}}f_2^{(11)}+(u_1-w_1)f^{(12)}\nonumber\\
\tilde{f}^{(21)}&=&-\tilde{v}_{\frac{3}{2}}f_2^{(11)}+(u_1-w_1)f^{(21)}
\end{eqnarray}
The problematic Massey products, we have encountered at orders $4$ and $5$ in deformation theory are of a similar character. Summing up the contributions, we find:
\begin{equation}
\bar{{f}}^{(12)}\cdot\mat{1}{0}{0}{0}\qquad \bar{{f}}^{(21)}\cdot\mat{0}{0}{0}{1},
\end{equation}
where
\begin{eqnarray}
\bar{f}^{(12)}&=&-\frac{1}{2}u_2\tilde{u}_{\frac{3}{2}}w_1^2+u_1\tilde{u}^2_{\frac{3}{2}}\tilde{w}_{\frac{3}{2}}-\frac{1}{2}\tilde{u}^2_{\frac{3}{2}}\tilde{w}_{\frac{3}{2}}w_1+2u_1^2u_2\tilde{u}_{\frac{3}{2}}-(1-2a)u_1^4\tilde{u}_{\frac{3}{2}}-\frac{1}{2}u_1^3\tilde{u}_{\frac{3}{2}}w_1\nonumber\\
&&-\frac{1}{2}(1+a)u_1^2\tilde{u}_{\frac{3}{2}}w_1^2-\frac{1}{2}u_1\tilde{u}_{\frac{3}{2}}w_1^3+\frac{1}{2}\tilde{u}_{\frac{3}{2}}w_1^4\nonumber\\
\bar{f}^{(21)}&=&-\frac{1}{2}u_2\tilde{w}_{\frac{3}{2}}w_1^2+u_1\tilde{u}_{\frac{3}{2}}\tilde{w}^2_{\frac{3}{2}}-\frac{1}{2}\tilde{u}_{\frac{3}{2}}\tilde{w}^2_{\frac{3}{2}}w_1+2u_1^2u_2\tilde{w}_{\frac{3}{2}}-(1-2a)u_1^4\tilde{w}_{\frac{3}{2}}-\frac{1}{2}u_1^3\tilde{w}_{\frac{3}{2}}w_1\nonumber\\
&&-\frac{1}{2}(1+a)u_1^2\tilde{w}_{\frac{3}{2}}w_1^2-\frac{1}{2}u_1\tilde{w}_{\frac{3}{2}}w_1^3+\frac{1}{2}\tilde{w}_{\frac{3}{2}}w_1^4
\end{eqnarray}
As it should be, these equations are just combinations of the regular F--terms:
\begin{eqnarray}
\bar{f}^{(12)}&=&(u_1-\frac{1}{2}w_1)f^{(12)}\nonumber\\
\bar{f}^{(21)}&=&(u_1-\frac{1}{2}w_1)f^{(21)}
\end{eqnarray}
\subsection{Bulk deformations}
\label{sec-a3bulk}
We now consider the deformed Landau--Ginzburg superpotential:
\begin{equation}
W=x^5+s_5+s_4x+s_3x^2+s_2x^3=x^5+\sum_is_i\phi_i,
\end{equation}
where the $s_i$ are deformation parameters whose weights are indicated by their indices. The starting point of the deformation theory calculation are the undeformed matrix factorizations $Q_1$, $Q_2$ of $x^5$ as in (\ref{a4mf}).\\
We note that the $s_5$ deformation can be interpreted as a bosonic open string state for both branes and will therefore contribute to the F--terms $f^{(11)}_1$ and $f^{(22)}$. The linear deformation with parameter $s_4$ is in the bosonic cohomology for $Q_1$ but exact with respect to $Q_2$. Thus, we get a contribution to the F--term $f^{(11)}_2$ and a new $\alpha^{(22)}$. The quadratic and the cubic deformation are exact for both branes which yields two $\alpha^{(11)}$'s and two new $\alpha^{(22)}$'s. So, in total we get five new deformations on the boundary and the vector $\vm$ has $3$ new entries, one for each bulk deformation which is $Q$-- exact on at least one brane. We extend the vector $\vm$ we had in the case without bulk deformations such that the first three entries account for the closed string deformations. The vector $\vm=(1,0,0,0,0,0,0,0)$ is associated to the quadratic deformation $x^2\cdot\mathbbm{1}$ with parameter $s_3$, The cubic deformation $x^3\cdot\mathbbm{1}$ is labelled by $\vm=(0,1,0,0,0,0,0,0)$. Finally, $x\cdot\mathbbm{1}$, which is only exact on $Q_2$ corresponds to  $\vm=(0,0,1,0,0,0,0,0)$. Now we have to find $\alpha^{(AA)}_{\vm}$\footnote{For this example the bulk deformations are neither physical nor exact in the boundary changing sector. If we were looking at a system of two branes where $Q_1=Q_2$ we also would get contributions $\alpha^{(AB)}_{\vm}$  in the off diagonal blocks.} at order $|\vm|=1$ such that $\{Q_A,\alpha^{(AA)}_{\vm}\}=\phi\cdot\mathbbm{1}$. This is the most general choice:
\begin{eqnarray}
\alpha^{(11)}_{(1,0,0,0,0,0,0,0)}=\mat{0}{0}{1}{0}&\quad&\alpha^{(22)}_{(1,0,0,0,0,0,0,0)}=\mat{0}{0}{x}{0}\nonumber\\
\alpha^{(11)}_{(0,1,0,0,0,0,0,0)}=\mat{0}{1-b}{bx}{0}&\quad&\alpha^{(22)}_{(0,1,0,0,0,0,0,0)}=\mat{0}{0}{x^2}{0}\nonumber\\
&\quad&\alpha^{(22)}_{(0,0,1,0,0,0,0,0)}=\mat{0}{0}{1}{0}
\end{eqnarray}
Here $b$ parametrizes an ambiguity in the choice of $\alpha^{(11)}_{(0,1,0,0,0,0,0,0)}$. The $\alpha_{\vm}$ with non--zero entries in the last five slots of $\vm$ coincide with those for the case without bulk deformations (\ref{a3lin}). Furthermore, we have to write the bulk deformations which are also bosonic open string states into the F--term equations:
\begin{eqnarray}
f^{1,(11)}_1:&\quad&s_5\nonumber\\
f^{1,(11)}_2:&\quad&s_4\nonumber\\
f^{1,(12)}:&\quad&0\nonumber\\
f^{1,(21)}:&\quad&0\nonumber\\
f^{1,(22)}:&\quad&s_5
\end{eqnarray} 
Note that these terms have to be added by hand and that the deformation theory algorithm is blind to these linear obstructions in the bulk parameters since they cannot be associated to a vector $\vm$. Therefore one cannot reduce the dimensions of the bases $B$, $\bar{B}$ by expressing higher order contributions to the F--terms in terms of these linear ones. \\
Now we are all set for computing higher order deformations. The algorithm now works just like in the case without bulk moduli, the only difference being the extended vectors $\vm$ and a different choice of bases $B$, $\bar{B}$. The details of the calculation are not very enlightening, so we only state the final result for the F--terms:
\begin{eqnarray}
\label{bulk-fterms}
f^{(11)}_1:&\quad&s_5+(1-b)s_2s_3+s_3u_2+2u_2\tilde{u}_{\frac{3}{2}}\tilde{w}_{\frac{3}{2}}+(1-2b)s_2\tilde{u}_{\frac{3}{2}}\tilde{w}_{\frac{3}{2}}+s_2u_1u_2+(2+a)s_3u_1^2\nonumber\\
&\quad&+(1-b)s_2^2u_1-\tilde{u}_{\frac{3}{2}}\tilde{w}_{\frac{3}{2}}w_1^2+u_1\tilde{u}_{\frac{3}{2}}\tilde{w}_{\frac{3}{2}}w_1+(3+2a)u_1^2\tilde{u}_{\frac{3}{2}}\tilde{w}_{\frac{3}{2}}+5u_1^3u_2\nonumber\\
&\quad&+(3+a-5b)s_2u_1^3+(-2+5a)u_1^5\nonumber\\
f^{(11)}_2:&\quad&s_4+s_3u_1+b(1-b)s_2^2-(1-2a)s_2u_2-u_2^2+\tilde{u}_{\frac{3}{2}}\tilde{w}_{\frac{3}{2}}w_1+u_1\tilde{u}_{\frac{3}{2}}\tilde{w}_{\frac{3}{2}}+(3-2a)u_1^2u_2\nonumber\\
&\quad&+(2-3b+a(-1+2b))s_2u_1^2-(1-3a+a^2)u_1^4\nonumber\\
f^{(12)}:&\quad&s_3\tilde{u}_{\frac{3}{2}}+\tilde{u}_{\frac{3}{2}}^2\tilde{w}_{\frac{3}{2}}+u_2\tilde{u}_{\frac{3}{2}}w_1+2u_1u_2\tilde{u}_{\frac{3}{2}}-bs_2\tilde{u}_{\frac{3}{2}}w_1+(1-2b)s_2u_1\tilde{u}_{\frac{3}{2}}-\tilde{u}_{\frac{3}{2}}w_1^3-u_1\tilde{u}_{\frac{3}{2}}w_1^2\nonumber\\
&\quad&-(1-a)u_1^2\tilde{u}_{\frac{3}{2}}w_1+(1-2a)u_1^3\tilde{u}_{\frac{3}{2}}\nonumber\\
f^{(21)}:&\quad&s_3\tilde{w}_{\frac{3}{2}}+\tilde{u}_{\frac{3}{2}}\tilde{w}_{\frac{3}{2}}^2+u_2\tilde{w}_{\frac{3}{2}}w_1+2u_1u_2\tilde{w}_{\frac{3}{2}}-bs_2\tilde{w}_{\frac{3}{2}}w_1+(1-2b)s_2u_1\tilde{w}_{\frac{3}{2}}-\tilde{w}_{\frac{3}{2}}w_1^3-u_1\tilde{w}_{\frac{3}{2}}w_1^2\nonumber\\
&\quad&-(1-a)u_1^2\tilde{w}_{\frac{3}{2}}w_1+(1-2a)u_1^3\tilde{w}_{\frac{3}{2}}\nonumber\\
f^{(22)}:&\quad&s_5+s_4w_1+u_2\tilde{u}_{\frac{3}{2}}\tilde{w}_{\frac{3}{2}}-as_2\tilde{u}_{\frac{3}{2}}\tilde{w}_{\frac{3}{2}}-s_3w_1^2-3\tilde{u}_{\frac{3}{2}}\tilde{w}_{\frac{3}{2}}w_1^2-2u_1\tilde{u}_{\frac{3}{2}}\tilde{w}_{\frac{3}{2}}w_1-(1-a)u_1^2\tilde{u}_{\frac{3}{2}}\tilde{w}_{\frac{3}{2}}\nonumber\\
&\quad&+s_2w_1^3+w_1^5
\end{eqnarray}
The parameter $a$ is the same as for the case without bulk moduli. In order to be able to integrate these F--terms one has to fix the parameter $a$ to the value $a=\frac{1}{2}$. This corresponds to a more symmetric choice of the deformation $\alpha^{(11)}_{(0,0,0,0,2,0,0,0)}$. Similar phenomena have also been encountered for two--parameter Calabi--Yau threefolds \cite{Knapp:2008tv}.\\
Fixing $a$, we find the following expression for the effective superpotential:
\begin{eqnarray}
\mathcal{W}_{eff}&=&\frac{1}{4}u_1^4u_2+u_1^2u_2^2-\frac{1}{3}u_2^3+2u_1u_2\tilde{u}_{\frac{3}{2}}\tilde{w}_{\frac{3}{2}}+\frac{1}{2}\tilde{u}_{\frac{3}{2}}^2\tilde{w}_{\frac{3}{2}}^2-\frac{1}{2}u_1^2\tilde{u}_{\frac{3}{2}}\tilde{w}_{\frac{3}{2}}w_1+u_2\tilde{u}_{\frac{3}{2}}\tilde{w}_{\frac{3}{2}}w_1\nonumber\\
&&-u_1\tilde{u}_{\frac{3}{2}}\tilde{w}_{\frac{3}{2}}w_1^2-\tilde{u}_{\frac{3}{2}}\tilde{w}_{\frac{3}{2}}w_1^3+\frac{1}{6}w_1^6\nonumber\\
&&+s_2\left(\frac{1-2b}{8}u_1^4+\frac{3-4b}{2}u_1^2u_2-\frac{1-2b}{2}u_2^2+(1-2b)u_1\tilde{u}_{\frac{3}{2}}\tilde{w}_{\frac{3}{2}}-b\tilde{u}_{\frac{3}{2}}\tilde{w}_{\frac{3}{2}}w_1+\frac{1}{4}w_1^4\right)\nonumber\\
&&+s_3\left(\frac{1}{6}u_1^3+u_1u_2+\tilde{u}_{\frac{3}{2}}\tilde{w}_{\frac{3}{2}}-\frac{1}{3}w_1^3\right)+s_2^2\left(\frac{1-3b+b^2}{2}u_1^2+b(1-b)u_2\right)\nonumber\\
&&+s_4\left(-u_1^2+u_2+\frac{1}{2}w_1^2\right)+s_5(u_1+w_1)+s_2s_3(1-b)u_1
\end{eqnarray}
The effective superpotential for this configuration has also been computed in \cite{Knapp:2006rd} by solving consistency constraints of disk amplitudes. One can check that the two results are {\em not} related through a field redefinition. The consistency constraints of disk amplitudes include in particular a generalized Cardy condition, which puts very strong restrictions on the terms in the effective superpotential. Without a certain truncation (cf. appendix B of \cite{Knapp:2007vc}) of the Cardy conditions in the boundary changing sector the consistency constraints do not have a solution at all. In the deformation theory calculation we had to fix the parameter $a$. This fixes a particular choice of field redefinition which does not seem to be compatible with the constraints from the Cardy condition.\\
Note that, as in the case without bulk deformations, the Massey products also give extra terms which look like obstructions. They do not contain new information and turn out to be homogeneous combinations of the equations (\ref{bulk-fterms}). Similarly $Q_{def}^2$ squares to the bulk deformed Landau--Ginzburg superpotential modulo  (\ref{bulk-fterms}). 
\subsection{Three branes}
\begin{figure}
\begin{center}
\includegraphics{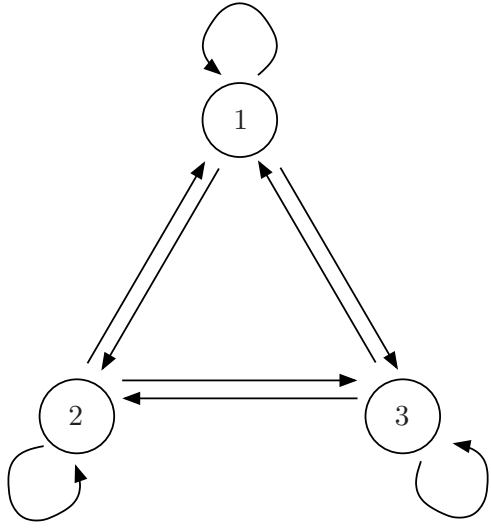}
\caption{The quiver diagram for the $A_4$ model with three branes.}\label{a3quiver3branes-fig}
\end{center}
\end{figure}

Just for amusement, we also compute an example with three branes. We choose:
\begin{equation}
Q_i=\mat{0}{x}{x^4}{0}\qquad i=1,2,3
\end{equation}
Due to the symmetry of the configuration, all the deformations and obstructions look the same:
\begin{equation}
\Psi^{(ij)}=\mat{0}{1}{-x^3}{0}\quad\Phi^{(ij)}=\mat{1}{0}{0}{1} \qquad i,j=1,2,3
\end{equation}
All fermions have R--charge $\frac{3}{5}$, all bosons have charge $0$. To each fermionic open string state, we associate a deformation parameter $u_{(ij)}$ which has weight $1$. The quiver diagram associated to this configuration is depicted in figure \ref{a3quiver3branes-fig}.
A calculation of the F--terms by hand would be torturous, so we use the {\tt mod\_versal} routine of the computer algebra program Singular \cite{GPS} to do the job for us. The F--term equations would fill several pages, so we do not give them here. The effective superpotential is a fairly large expression:

{\small
\begin{eqnarray}
\mathcal{W}_{eff}&=&\frac{1}{6} u_{(11)}^6+u_{(12)} u_{(21)} u_{(11)}^4+u_{(13)} u_{(31)} u_{(11)}^4+u_{(12)} u_{(21)} u_{(22)} u_{(11)}^3+u_{(12)} u_{(23)}
   u_{(31)} u_{(11)}^3\nonumber\\
&&+u_{(13)} u_{(21)} u_{(32)} u_{(11)}^3+u_{(13)} u_{(31)} u_{(33)} u_{(11)}^3+\frac{3}{2} u_{(12)}^2 u_{(21)}^2
   u_{(11)}^2+u_{(12)} u_{(21)} u_{(22)}^2 u_{(11)}^2\nonumber\\
&&+\frac{3}{2} u_{(13)}^2 u_{(31)}^2 u_{(11)}^2+u_{(13)} u_{(31)} u_{(33)}^2
   u_{(11)}^2+3 u_{(12)} u_{(13)} u_{(21)} u_{(31)} u_{(11)}^2\nonumber\\
&&+u_{(12)} u_{(22)} u_{(23)} u_{(31)} u_{(11)}^2+u_{(13)} u_{(21)} u_{(22)}
   u_{(32)} u_{(11)}^2+u_{(12)} u_{(21)} u_{(23)} u_{(32)} u_{(11)}^2\nonumber\\
&&+u_{(13)} u_{(23)} u_{(31)} u_{(32)} u_{(11)}^2+u_{(12)} u_{(23)}
   u_{(31)} u_{(33)} u_{(11)}^2+u_{(13)} u_{(21)} u_{(32)} u_{(33)} u_{(11)}^2\nonumber\\
&&+u_{(12)} u_{(21)} u_{(22)}^3 u_{(11)}+u_{(13)} u_{(31)}
   u_{(33)}^3 u_{(11)}+2 u_{(12)} u_{(13)} u_{(23)} u_{(31)}^2 u_{(11)}+u_{(13)} u_{(21)} u_{(23)} u_{(32)}^2 u_{(11)}\nonumber\\
&&+u_{(12)} u_{(23)}
   u_{(31)} u_{(33)}^2 u_{(11)}+u_{(13)} u_{(21)} u_{(32)} u_{(33)}^2 u_{(11)}+2 u_{(12)}^2 u_{(21)}^2 u_{(22)} u_{(11)}\nonumber\\
&&+2 u_{(12)}
   u_{(13)} u_{(21)} u_{(22)} u_{(31)} u_{(11)}+u_{(12)} u_{(22)}^2 u_{(23)} u_{(31)} u_{(11)}+2 u_{(12)}^2 u_{(21)} u_{(23)} u_{(31)}
   u_{(11)}\nonumber\\
&&+2 u_{(12)} u_{(13)} u_{(21)}^2 u_{(32)} u_{(11)}+u_{(13)} u_{(21)} u_{(22)}^2 u_{(32)} u_{(11)}+2 u_{(12)} u_{(21)} u_{(22)}
   u_{(23)} u_{(32)} u_{(11)}\nonumber\\
&&+u_{(12)} u_{(23)}^2 u_{(31)} u_{(32)} u_{(11)}+2 u_{(13)}^2 u_{(21)} u_{(31)} u_{(32)} u_{(11)}+u_{(13)}
   u_{(22)} u_{(23)} u_{(31)} u_{(32)} u_{(11)}\nonumber\\
&&+2 u_{(13)}^2 u_{(31)}^2 u_{(33)} u_{(11)}+2 u_{(12)} u_{(13)} u_{(21)} u_{(31)} u_{(33)}
   u_{(11)}+u_{(12)} u_{(22)} u_{(23)} u_{(31)} u_{(33)} u_{(11)}\nonumber\\
&&+u_{(13)} u_{(21)} u_{(22)} u_{(32)} u_{(33)} u_{(11)}+u_{(12)} u_{(21)}
   u_{(23)} u_{(32)} u_{(33)} u_{(11)}+2 u_{(13)} u_{(23)} u_{(31)} u_{(32)} u_{(33)} u_{(11)}\nonumber\\
&&+\frac{1}{6} u_{(22)}^6+\frac{1}{6}
   u_{(33)}^6+u_{(12)} u_{(21)} u_{(22)}^4+u_{(13)} u_{(31)} u_{(33)}^4+u_{(23)} u_{(32)} u_{(33)}^4+\frac{1}{3} u_{(12)}^3
   u_{(21)}^3+\frac{1}{3} u_{(13)}^3 u_{(31)}^3\nonumber\\
&&+\frac{1}{3} u_{(23)}^3 u_{(32)}^3+u_{(12)} u_{(23)} u_{(31)} u_{(33)}^3+u_{(13)}
   u_{(21)} u_{(32)} u_{(33)}^3+u_{(22)} u_{(23)} u_{(32)} u_{(33)}^3+\frac{3}{2} u_{(12)}^2 u_{(21)}^2 u_{(22)}^2\nonumber\\
&&+\frac{1}{2}
   u_{(12)}^2 u_{(23)}^2 u_{(31)}^2+u_{(12)} u_{(13)}^2 u_{(21)} u_{(31)}^2+u_{(12)} u_{(13)} u_{(22)} u_{(23)} u_{(31)}^2+\frac{1}{2}
   u_{(13)}^2 u_{(21)}^2 u_{(32)}^2\nonumber\\
&&+\frac{3}{2} u_{(22)}^2 u_{(23)}^2 u_{(32)}^2+u_{(12)} u_{(21)} u_{(23)}^2 u_{(32)}^2+2 u_{(13)}
   u_{(21)} u_{(22)} u_{(23)} u_{(32)}^2+u_{(13)} u_{(23)}^2 u_{(31)} u_{(32)}^2\nonumber\\
&&+\frac{3}{2} u_{(13)}^2 u_{(31)}^2
   u_{(33)}^2+\frac{3}{2} u_{(23)}^2 u_{(32)}^2 u_{(33)}^2+u_{(12)} u_{(13)} u_{(21)} u_{(31)} u_{(33)}^2+u_{(12)} u_{(22)} u_{(23)}
   u_{(31)} u_{(33)}^2\nonumber\\
&&+u_{(13)} u_{(21)} u_{(22)} u_{(32)} u_{(33)}^2+u_{(22)}^2 u_{(23)} u_{(32)} u_{(33)}^2+u_{(12)} u_{(21)} u_{(23)}
   u_{(32)} u_{(33)}^2\nonumber\\
&&+3 u_{(13)} u_{(23)} u_{(31)} u_{(32)} u_{(33)}^2+u_{(12)}^2 u_{(13)} u_{(21)}^2 u_{(31)}+u_{(12)} u_{(13)} u_{(21)}
   u_{(22)}^2 u_{(31)}+u_{(12)} u_{(22)}^3 u_{(23)} u_{(31)}\nonumber\\
&&+2 u_{(12)}^2 u_{(21)} u_{(22)} u_{(23)} u_{(31)}+u_{(13)} u_{(21)} u_{(22)}^3
   u_{(32)}+u_{(13)}^2 u_{(23)} u_{(31)}^2 u_{(32)}+2 u_{(12)} u_{(13)} u_{(21)}^2 u_{(22)} u_{(32)}\nonumber\\
&&+u_{(22)}^4 u_{(23)}
   u_{(32)}+u_{(12)}^2 u_{(21)}^2 u_{(23)} u_{(32)}+3 u_{(12)} u_{(21)} u_{(22)}^2 u_{(23)} u_{(32)}+2 u_{(12)} u_{(22)} u_{(23)}^2
   u_{(31)} u_{(32)}\nonumber\\
&&+u_{(13)}^2 u_{(21)} u_{(22)} u_{(31)} u_{(32)}+u_{(13)} u_{(22)}^2 u_{(23)} u_{(31)} u_{(32)}+3 u_{(12)} u_{(13)}
   u_{(21)} u_{(23)} u_{(31)} u_{(32)}\nonumber\\
&&+2 u_{(12)} u_{(13)} u_{(23)} u_{(31)}^2 u_{(33)}+2 u_{(22)} u_{(23)}^2 u_{(32)}^2 u_{(33)}+2
   u_{(13)} u_{(21)} u_{(23)} u_{(32)}^2 u_{(33)}\nonumber\\
&&+u_{(12)} u_{(13)} u_{(21)} u_{(22)} u_{(31)} u_{(33)}+u_{(12)} u_{(22)}^2 u_{(23)} u_{(31)}
   u_{(33)}+u_{(12)}^2 u_{(21)} u_{(23)} u_{(31)} u_{(33)}\nonumber\\
&&+u_{(12)} u_{(13)} u_{(21)}^2 u_{(32)} u_{(33)}+u_{(13)} u_{(21)} u_{(22)}^2
   u_{(32)} u_{(33)}+u_{(22)}^3 u_{(23)} u_{(32)} u_{(33)}\nonumber\\
&&+2 u_{(12)} u_{(21)} u_{(22)} u_{(23)} u_{(32)} u_{(33)}+2 u_{(12)} u_{(23)}^2
   u_{(31)} u_{(32)} u_{(33)}+2 u_{(13)}^2 u_{(21)} u_{(31)} u_{(32)} u_{(33)}\nonumber\\
&&+2 u_{(13)} u_{(22)} u_{(23)} u_{(31)} u_{(32)} u_{(33)}
\end{eqnarray}
}
\section{The $E_6$ minimal model}
\label{sec-e6}
In order to show that the procedure does not only work for the simplest minimal model we now discuss a system of two D--branes in the $E_6$ minimal model.
\subsection{Setup}
\begin{figure}
\begin{center}
\includegraphics{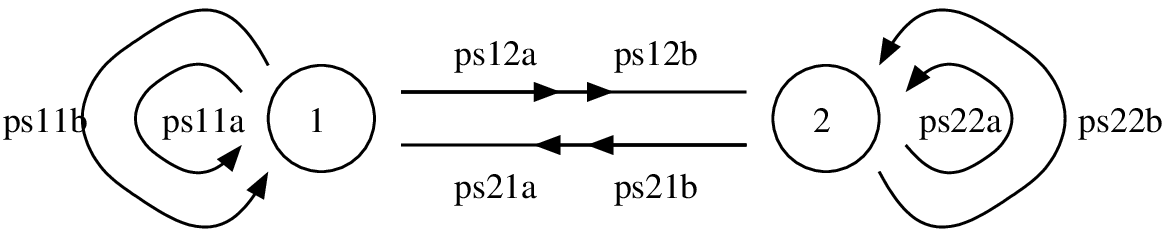}
\caption{The quiver diagram for the $E_6$ model with two branes.}\label{e6quiver-fig}
\end{center}
\end{figure}

The Landau--Ginzburg superpotential of the $E_6$ minimal model is:
\begin{equation}
W=x^3+y^4-z^2
\end{equation}
 We consider a pair of matrix factorizations $Q_i=\mat{0}{E_i}{J_i}{0}$, $i=1,2$ with:
\begin{equation}
E_1=J_2=\mat{-y^2-z}{x}{x^2}{y^2-z}\qquad J_1=E_2=\mat{-y^2+z}{x}{x^2}{y^2+z}
\end{equation}
Note that this is a brane--antibrane pair. The boundary preserving fermions have the same shape on both branes:
\begin{equation}
\Psi^{(11)}_1=\Psi^{(22)}_1=\left(\begin{array}{cccc}
0&0&0&1\\
0&0&-x&0\\
0&1&0&0\\
-x&0&0&0
\end{array}\right)\qquad
\Psi^{(11)}_2=\Psi^{(22)}_2=\left(\begin{array}{cccc}
0&0&0&y\\
0&0&-xy&0\\
0&y&0&0\\
-xy&0&0&0
\end{array}\right)
\end{equation}
These states have R--charge $\frac{1}{3}$ and $\frac{5}{6}$, respectively. The bosonic open string states in the boundary preserving sector look as follows:
\begin{equation}
\Phi^{(11)}_1=\Phi^{(22)}_1=\left(\begin{array}{cccc}
1&0&0&0\\
0&1&0&0\\
0&0&1&0\\
0&0&0&1
\end{array}\right)\qquad
\Phi^{(11)}_2=\Phi^{(22)}_2=\left(\begin{array}{cccc}
y&0&0&0\\
0&y&0&0\\
0&0&y&0\\
0&0&0&y
\end{array}\right)
\end{equation}
These have charges $0$ and $\frac{1}{2}$. Via Serre duality, they pair up with the fermions such that the sum of charges is $\frac{5}{6}$. In the boundary changing sector, the fermionic open string states have charges $0$ and $\frac{1}{2}$:
\begin{equation}
\Psi^{(12)}_1=\Psi^{(21)}_1=\left(\begin{array}{cccc}
0&0&1&0\\
0&0&0&1\\
-1&0&0&0\\
0&-1&0&0
\end{array}\right)\qquad
\Psi^{(12)}_2=\Psi^{(21)}_2=\left(\begin{array}{cccc}
0&0&y&0\\
0&0&0&y\\
-y&0&0&0\\
0&-y&0&0
\end{array}\right)
\end{equation}
Their bosonic partners have charges $\frac{1}{3}$ and $\frac{5}{6}$:
\begin{equation}
\Phi^{(12)}_1=\Phi^{(21)}_1=\left(\begin{array}{cccc}
0&1&0&0\\
-x&0&0&0\\
0&0&0&-1\\
0&0&x&0
\end{array}\right)\qquad
\Phi^{(12)}_2=\Phi^{(21)}_2=\left(\begin{array}{cccc}
0&y&0&0\\
-xy&0&0&0\\
0&0&0&-y\\
0&0&xy&0
\end{array}\right)
\end{equation}
The quiver diagram is depicted in figure \ref{e6quiver-fig}.
We associate deformation parameters labelled by their weights to the fermionic open string states. The linear deformation of the matrix factorization has the following form:
\begin{equation}
Q_{def}^{lin}=\mat{Q_1+u_4\Psi^{(11)}_1+u_1\Psi^{(11)}_2}{\tilde{u}_{6}\Psi^{(12)}_1+\tilde{u}_{3}\Psi^{(12)}_2}{\tilde{w}_{6}\Psi^{(21)}_1+\tilde{w}_{3}\Psi^{(21)}_2}{Q_2+w_4\Psi^{(22)}_1+w_1\Psi^{(22)}_2}
\end{equation}
\subsection{Deformation Theory}
The deformation theory algorithm is already too tedious to do by hand, even for this configuration, which is the simplest for $E_6$. With the help of Singular we find the following F--term equations:
\begin{eqnarray}
f^{(11)}_1:&\quad&-\tilde{u}_6\tilde{w}_6-u_4^3+\frac{1}{4}\tilde{u}_3^2\tilde{w}_3^2+2\tilde{u}_3\tilde{w}_3u_1^2u_4-\tilde{u}_3\tilde{w}_3u_1u_4w_1+\frac{1}{2}\tilde{u}_3\tilde{w}_3u_4w_1^2+\frac{1}{4}\tilde{u}_3\tilde{w}_3u_1^2w_4\nonumber\\
&\quad&+\frac{1}{2}\tilde{u}_3\tilde{w}_3u_1w_1w_4+\frac{1}{4}\tilde{u}_3\tilde{w}_3w_1^2w_4+\frac{9}{4}u_1^4u_4^2-\frac{1}{8}\tilde{u}_3\tilde{w}_3u_1^6-\frac{3}{8}u_1^8u_4+\frac{1}{64}u_1^{12}\nonumber\\
f^{(11)}_2:&\quad&-\tilde{u}_6\tilde{w}_3-\tilde{u}_3\tilde{w}_6-3u_1u_4^2+\tilde{u}_3\tilde{w}_3u_1^3-\frac{3}{4}\tilde{u}_3\tilde{w}_3u_1^2w_1+\frac{1}{4}\tilde{u}_3\tilde{w}_3w_1^3+\frac{3}{2}u_1^5u_4-\frac{1}{8}u_1^9\nonumber\\
f^{(12)}_1:&\quad&-\tilde{u}_6u_1-\tilde{u}_3u_4+\tilde{u}_6w_1+\tilde{u}_3w_4+\frac{1}{4}\tilde{u}_3u_1^4\frac{1}{4}\tilde{u}_3u_1^3w_1+\frac{1}{4}\tilde{u}_3u_1w_1^3-\frac{1}{4}\tilde{u}_3w_1^4\nonumber\\
f^{(12)}_2:&\quad&-\tilde{u}_6u_4+\tilde{u}_6w_4+\frac{1}{2}\tilde{u}_3^2\tilde{w}_3u_1-\frac{1}{2}\tilde{u}_3^2\tilde{w}_3w_1+\frac{3}{4}\tilde{u}_3u_1^3u_4-\frac{3}{4}\tilde{u}_3u_1^2u_4w_1+\frac{3}{4}\tilde{u}_3u_1w_1^2w_4\nonumber\\
&\quad&-\frac{3}{4}\tilde{u}_3w_1^3w_4-\frac{1}{16}\tilde{u}_3u_1^7+\frac{1}{16}\tilde{u}_3u_1^6w_1-\frac{1}{16}\tilde{u}_3u_1w_1^6+\frac{1}{16}\tilde{u}_3w_1^7\nonumber\\
f^{(21)}_1:&\quad&\tilde{w}_6u_1+\tilde{w}_3u_4-\tilde{w}_6w_1-\tilde{w}_3w_4-\frac{1}{4}\tilde{w}_3u_1^4+\frac{1}{4}\tilde{w}_3u_1^3w_1-\frac{1}{4}\tilde{w}_3u_1w_1^3+\frac{1}{4}\tilde{w}_3w_1^4\nonumber\\
f^{(21)}_2:&\quad&\tilde{w}_6u_4-\tilde{w}_6w_4-\frac{1}{2}\tilde{u}_3\tilde{w}_3^2u_1+\frac{1}{2}\tilde{u}_3\tilde{w}_3^2w_1-\frac{3}{4}\tilde{w}_3u_1^3u_4+\frac{3}{4}\tilde{w}_3u_1^2u_4w_1-\frac{3}{4}\tilde{w}_3u_1w_1^2w_4\nonumber\\
&\quad&+\frac{3}{4}\tilde{w}_3w_1^3w_4+\frac{1}{16}\tilde{w}_3u_1^7-\frac{1}{16}\tilde{w}_3u_1^6w_1+\frac{1}{16}\tilde{w}_3u_1w_1^6-\frac{1}{16}\tilde{w}_3w_1^7\nonumber\\
f^{(22)}_1:&\quad&-\tilde{u}_6\tilde{w}_6-w_4^3+\frac{1}{4}\tilde{u}_3^2\tilde{w}_3^2+\frac{1}{4}\tilde{u}_3\tilde{w}_3u_1^2u_4-\frac{1}{2}\tilde{u}_3\tilde{w}_3u_1u_4w_1+\frac{1}{4}\tilde{u}_3\tilde{w}_3u_4w_1^2+\frac{1}{2}\tilde{u}_3\tilde{w}_3u_1^2w_4\nonumber\\
&\quad&-\tilde{u}_3\tilde{w}_3u_1w_1w_4+2\tilde{u}_3\tilde{w}_3w_1^2w_4+\frac{9}{4}w_1^4w_4^2-\frac{1}{8}\tilde{u}_3\tilde{w}_3w_1^6-\frac{3}{8}w_1^8w_4+\frac{1}{64}w_1^{12}\nonumber\\
f^{(22)}_2:&\quad&-\tilde{u}_6\tilde{w}_3-\tilde{u}_3\tilde{w}_6-3w_1w_4^2+\frac{1}{4}\tilde{u}_3\tilde{w}_3u_1^3-\frac{3}{4}\tilde{u}_3\tilde{w}_3u_1w_1^2 + \tilde{u}_3\tilde{w}_3w_1^3 + \frac{3}{2}w_1^5w_ -\frac{1}{8} w_1^9
\end{eqnarray}
The integration of these equations is quite involved. The effective superpotential has weight $13$. Our F--term equations have weights $\{12,9,7,10,7,10,12,9\}$, respectively. The eight integration variables have weights $\{1,3,4,6\}$, where we always have two variables with the same weights. In order to integrate correctly, we have to find homogeneous combinations of the F--terms which have weights such that we get weight $13$ when we add the weight of the integration variable. The integration with respect to the weight $1$ variables will be the most complicated. For that, we have to build a weight $12$ polynomial out of the F--terms. The ansatz for this looks as follows:
\begin{equation}
p_0f^{(11)}_1+p'_0f^{(22)}_1+p_3f^{(12)}_2+p'_3f^{(21)}_2+p_4f^{(11)}_2+p'_4f^{(22)}_2+p_5f^{(12)}_1+p'_5f^{(21)}_1,
\end{equation} 
where the $p_i,p'_i$ are homogeneous polynomials of degree $i$ in the deformation parameters with arbitrary numerical coefficients. We make such an ansatz for every integration variable and determine the unknown coefficients by the conditions that second order derivatives of the integral have to match pairwise. Up to an overall constant, we find the following expression for the effective superpotential:
\begin{eqnarray}
\mathcal{W}_{eff}&=&\frac{5}{832}u_1^{13}-\frac{1}{8}u_1^9u_4+\frac{3}{4}u_1^5u_4^2-u_1u_4^3+\frac{1}{4}\tilde{w}_6u_1^4\tilde{u}_3+\frac{1}{8}\tilde{w}_3u_1^7\tilde{u}_3+\tilde{w}_3u_1^3u_4\tilde{u}_3\nonumber\\
&&+\frac{1}{4}\tilde{w}_3^2u_1\tilde{u}_3^2-\tilde{w}_6u_1\tilde{u}_6+\frac{1}{4}\tilde{w}_3u_1^4\tilde{u}_6-\frac{1}{4}\tilde{w}_6u_1^3\tilde{u}_3w_1+\frac{1}{8}\tilde{w}_3u_1^6\tilde{u}_3w_1-\frac{3}{4}\tilde{w}_3u_1^2u_4\tilde{u}_3w_1\nonumber\\
&&-\frac{1}{4}\tilde{w}_3u_1^3\tilde{u}_6w_1+\frac{1}{4}\tilde{w}_6u_1\tilde{u}_3w_1^3-\frac{1}{8}\tilde{w}_3u_1^4\tilde{u}_3w_1^3+\frac{1}{4}\tilde{w}_3u_1\tilde{u}_6w_1^3+\frac{1}{8}\tilde{w}_3u_1^3\tilde{u}_3w_1^4\nonumber\\
&&-\frac{1}{8}\tilde{w}_3u_1\tilde{u}_3w_1^6-\frac{1}{4}\tilde{w}_3u_1^3\tilde{u}_3w_4+\frac{3}{4}\tilde{w}_3u_1\tilde{u}_3w_1^2w_4-\frac{1}{4}\tilde{w}_3^2\tilde{u}_3^2w_1+\tilde{w}_6\tilde{u}_6w_1+\frac{1}{4}\tilde{w}_3u_4\tilde{u}_3w_1^3\nonumber\\
&&-\frac{1}{4}\tilde{w}_6\tilde{u}_3w_1^4-\frac{1}{4}\tilde{w}_3\tilde{u}_6w_1^4+\frac{1}{8}\tilde{w}_3\tilde{u}_3w_1^7-\frac{5}{832}w_1^{13}-\tilde{w}_3\tilde{u}_3w_1^3w_4+\frac{1}{8}w_1^9w_4-\frac{3}{4}w_1^5w_4^2\nonumber\\
&&+w_1w_4^3-\tilde{w}_6u_4\tilde{u}_3-\tilde{w}_3u_4u_6+\tilde{w}_3\tilde{u}_6w_4+\tilde{w}_6\tilde{u}_3w_4
\end{eqnarray}
\section{Comments on the Calabi--Yau case}
\label{sec-cy}
In \cite{Knapp:2008tv} it was shown that the Massey product algorithm can also be applied to Calabi--Yau threefolds without any changes. The reason is that in complex dimension $3$ (and only there) the brane moduli are Serre dual to the obstructions. Therefore deformations of D--branes in Calabi--Yau threefolds are generically obstructed. This gives rise to an effective superpotential and a domain wall tension which, on the mirror, is the generating function of open Gromov--Witten invariants. Due to the presence of obstructed moduli Calabi--Yau threefolds have a lot in common with minimal models, at least as far as deformation theory is concerned. As a bonus, we also found in \cite{Knapp:2008tv} that in many examples also unobstructed moduli only lead to a finite number of deformations of the matrix factorization. This means that the deformation theory algorithm terminates after a finite number of steps even though it would not have to because we are dealing with true moduli. Given these nice features, the odds are actually good that the Massey product method is suited to compute effective superpotentials near Gepner points on Calabi--Yau threefolds also for models with more than one brane. The mirrors of such models are intersecting brane models which are of great interest from the phenomenological point of view. \\
Even if the deformation theory algorithm works out for more than one brane on a Calabi--Yau threefold the calculation is expected to be more challenging than for minimal models. In order to reproduce results for superpotentials on the quintic known in the literature (see for instance \cite{Brunner:1999jq,Douglas:2002fr,Ashok:2004xq,Baumgartl:2007an}) we will have to deal with more branes, more open string states and larger matrix factorizations. This will require some efficient computer code. Such a program must have the following features: The program has to compute the open string spectrum efficiently. In particular one should be able to work with multiply graded polynomial rings which implement orbifold actions. Only these orbifold invariant open string states should enter the deformation theory calculation. Furthermore, one has to include the additional labelling for the boundary changing sector, or, in other words, extend the algorithm to deformations theory of a system with more than one brane without getting gigantic matrices. 
\section{Conclusions}
\label{sec-conc}
In this paper we have extended the deformation theory algorithm of matrix factorizations to setups with more than one D--brane. The extension turned out to be somewhat trivial since one can make the formalism work like in a setup with just one D--brane by collecting the matrix factorizations of the component branes into one big matrix factorization. The reason why this works is due to the non--geometric and the categorical nature of matrix factorizations. In the matrix factorizations language it makes no difference if one has a single elementary brane or a bound state of a set of branes or a system of several D--branes. Of course the information whether one has one or more branes is encoded in the matrix factorizations, even if it may not always be easy to see. However, the point is that this additional structure does not make fundamental changes in the deformation theory algorithm. This line of argument also holds for the deformations, i.e. open string states. The only information about open string states we need is that they are physical in the sense of the cohomology of the matrix factorization. However, the formalism does not care about whether the deformations are in the boundary preserving or in the boundary changing sector. So formally, for the deformation theory of matrix factorizations it makes no difference\footnote{Of course, for exactly marginal deformations it may happen that the deformation theory algorithm does not terminate at a given order in deformation theory, but the rules of the algorithm do not change.} if we deform the brane with an open modulus, i.e. with a boundary preserving open string state, or if we switch on a tachyon between two D--branes, as it happens in the boundary changing sector.\\
Still, additional structure can be useful for practical purposes. Collecting many matrix factorizations into a big one to describe a system of multiple branes, one quickly ends up with huge matrix factorizations, which certainly leads to technical problems. In order to avoid this we have extended the Massey product algorithm by introducing an additional labelling which marks the starting end the end brane of a deformation. In this way we can work with the components of the big matrix factorizations and avoid computing products which are $0$ due to the block structure of the matrix factorization of the full system. Furthermore we discussed how to include bulk deformation into the extended deformation theory algorithm.\\
In order to demonstrate that the deformation theory works also for systems with more than one D--brane we have computed several examples for minimal models and reproduced a result which has been derived by other means in the literature. These examples are a realistic playground for models with more than one D--brane in Calabi--Yau threefolds. In the boundary preserving sector these have been shown to behave essentially like minimal models as far as deformation theory is concerned \cite{Knapp:2008tv}. The minimal model examples indicate that this may work equally well in the boundary changing sector. This gives a lot of material for further research \cite{wip}.

\bibliographystyle{fullsort}
\bibliography{massey}

\providecommand{\href}[2]{#2}\begingroup\raggedright\begin{thebibliography}{10}

\bibitem{Walcher:2006rs}
J.~Walcher, ``{Opening mirror symmetry on the quintic},'' {\em Commun. Math.
  Phys.} {\bf 276} (2007) 671--689,
\href{http://www.arXiv.org/abs/hep-th/0605162}{{\tt hep-th/0605162}}.

\bibitem{Morrison:2007bm}
D.~R. Morrison and J.~Walcher, ``{D-branes and Normal Functions},''
\href{http://www.arXiv.org/abs/arXiv:0709.4028[hep-th]}{{\tt
  arXiv:0709.4028[hep-th]}}.

\bibitem{Krefl:2008sj}
D.~Krefl and J.~Walcher, ``{Real Mirror Symmetry for One-parameter
  Hypersurfaces},'' {\em JHEP} {\bf 09} (2008) 031,
\href{http://www.arXiv.org/abs/arXiv:0805.0792[hep-th]}{{\tt
  arXiv:0805.0792[hep-th]}}.

\bibitem{Knapp:2008uw}
J.~Knapp and E.~Scheidegger, ``{Towards Open String Mirror Symmetry for
  One-Parameter Calabi-Yau Hypersurfaces},''
\href{http://www.arXiv.org/abs/arXiv:0805.1013[hep-th]}{{\tt
  arXiv:0805.1013[hep-th]}}.

\bibitem{Jockers:2008pe}
H.~Jockers and M.~Soroush, ``{Effective superpotentials for compact D5-brane
  Calabi-Yau geometries},'' {\em Commun. Math. Phys.} {\bf 290} (2009)
  249--290,
\href{http://www.arXiv.org/abs/arXiv:0808.0761[hep-th]}{{\tt
  arXiv:0808.0761[hep-th]}}.

\bibitem{Grimm:2008dq}
T.~W. Grimm, T.-W. Ha, A.~Klemm, and D.~Klevers, ``{The D5-brane effective
  action and superpotential in N=1 compactifications},'' {\em Nucl. Phys.} {\bf
  B816} (2009) 139--184,
\href{http://www.arXiv.org/abs/arXiv:0811.2996[hep-th]}{{\tt
  arXiv:0811.2996[hep-th]}}.

\bibitem{Alim:2009rf}
M.~Alim, M.~Hecht, P.~Mayr, and A.~Mertens, ``{Mirror Symmetry for Toric Branes
  on Compact Hypersurfaces},''
\href{http://www.arXiv.org/abs/arXiv:0901.2937[hep-th]}{{\tt
  arXiv:0901.2937[hep-th]}}.

\bibitem{Jockers:2009mn}
H.~Jockers and M.~Soroush, ``{Relative periods and open-string integer
  invariants for a compact Calabi-Yau hypersurface},''
\href{http://www.arXiv.org/abs/arXiv:0904.4674[hep-th]}{{\tt
  arXiv:0904.4674[hep-th]}}.

\bibitem{Walcher:2009uj}
J.~Walcher, ``{Calculations for Mirror Symmetry with D-branes},''
\href{http://www.arXiv.org/abs/arXiv:0904.4905[hep-th]}{{\tt
  arXiv:0904.4905[hep-th]}}.

\bibitem{Siqveland1}
A.~Siqveland, ``{T}he {M}ethod of {C}omputing {F}ormal {M}oduli,'' {\em J.
  Alg.} {\bf 241} (2001) 292--327.

\bibitem{Knapp:2006rd}
J.~Knapp and H.~Omer, ``{M}atrix {F}actorizations, {M}inimal {M}odels and
  {M}assey {P}roducts,'' {\em JHEP} {\bf 05} (2006) 064,
\href{http://www.arXiv.org/abs/hep-th/0604189}{{\tt hep-th/0604189}}.

\bibitem{Knapp:2008tv}
J.~Knapp and E.~Scheidegger, ``{Matrix Factorizations, Massey Products and
  F-Terms for Two-Parameter Calabi-Yau Hypersurfaces},''
\href{http://www.arXiv.org/abs/arXiv:0812.2429[hep-th]}{{\tt
  arXiv:0812.2429[hep-th]}}.

\bibitem{Hori:2004ja}
K.~Hori and J.~Walcher, ``{F-term equations near Gepner points},'' {\em JHEP}
  {\bf 01} (2005) 008,
\href{http://www.arXiv.org/abs/hep-th/0404196}{{\tt hep-th/0404196}}.

\bibitem{Knapp:2007vc}
J.~Knapp, ``{D-Branes in Topological String Theory},''
  \href{http://www.arXiv.org/abs/arXiv:0709.2045[hep-th]}{{\tt
  arXiv:0709.2045[hep-th]}}.
PhD thesis.

\bibitem{Jockers:2007ng}
H.~Jockers and W.~Lerche, ``{Matrix Factorizations, D-Branes and their
  Deformations},'' {\em Nucl. Phys. Proc. Suppl.} {\bf 171} (2007) 196--214,
\href{http://www.arXiv.org/abs/arXiv:0708.0157[hep-th]}{{\tt
  arXiv:0708.0157[hep-th]}}.

\bibitem{Herbst:2004jp}
M.~Herbst, C.-I. Lazaroiu, and W.~Lerche, ``{Superpotentials, A(infinity)
  relations and WDVV equations for open topological strings},'' {\em JHEP} {\bf
  02} (2005) 071,
\href{http://www.arXiv.org/abs/hep-th/0402110}{{\tt hep-th/0402110}}.

\bibitem{Herbst:2004zm}
M.~Herbst, C.-I. Lazaroiu, and W.~Lerche, ``{D-brane effective action and
  tachyon condensation in topological minimal models},'' {\em JHEP} {\bf 03}
  (2005) 078,
\href{http://www.arXiv.org/abs/hep-th/0405138}{{\tt hep-th/0405138}}.

\bibitem{GPS}
G.-M. Greuel, G.~Pfister, and H.~Sch{\"o}nemann, ``{\sc Singular} 3.1.0 --- {A}
  computer algebra system for polynomial computations,''.
  http://www.singular.uni-kl.de.

\bibitem{Brunner:1999jq}
I.~Brunner, M.~R. Douglas, A.~E. Lawrence, and C.~Romelsberger, ``{D-branes on
  the quintic},'' {\em JHEP} {\bf 08} (2000) 015,
\href{http://www.arXiv.org/abs/hep-th/9906200}{{\tt hep-th/9906200}}.

\bibitem{Douglas:2002fr}
M.~R. Douglas, S.~Govindarajan, T.~Jayaraman, and A.~Tomasiello, ``{D-branes on
  Calabi-Yau manifolds and superpotentials},'' {\em Commun. Math. Phys.} {\bf
  248} (2004) 85--118,
\href{http://www.arXiv.org/abs/hep-th/0203173}{{\tt hep-th/0203173}}.

\bibitem{Ashok:2004xq}
S.~K. Ashok, E.~Dell'Aquila, D.-E. Diaconescu, and B.~Florea, ``{Obstructed
  D-branes in Landau-Ginzburg orbifolds},'' {\em Adv. Theor. Math. Phys.} {\bf
  8} (2004) 427--472,
\href{http://www.arXiv.org/abs/hep-th/0404167}{{\tt hep-th/0404167}}.

\bibitem{Baumgartl:2007an}
M.~Baumgartl, I.~Brunner, and M.~R. Gaberdiel, ``{D-brane superpotentials and
  RG flows on the quintic},'' {\em JHEP} {\bf 07} (2007) 061,
\href{http://www.arXiv.org/abs/arXiv:0704.2666[hep-th]}{{\tt
  arXiv:0704.2666[hep-th]}}.

\bibitem{wip}
J.~Knapp and E.~Scheidegger, ``work in progress.''.

\end{thebibliography}\endgroup

\end{document}